\documentclass[twocolumn]{openjournal}
\usepackage{graphicx} 

\usepackage[colorlinks,linkcolor=blue,citecolor=blue,urlcolor=blue ]{hyperref}
\usepackage[utf8]{inputenc}
\usepackage{float}
\usepackage{xcolor}
\usepackage{ulem}
\usepackage[T1]{fontenc}

\makeatletter

\newcommand{\Rmnum}[1]{\expandafter\@slowromancap\romannumeral #1@}

\newcommand{\gsim}{\lower0.6ex\vbox{\hbox{$\buildrel{\textstyle >}\over{\sim}\ $}}}

\begin{document}

\title[short title]
{$z \sim 2$ dual AGN host galaxies are disky: stellar kinematics in the ASTRID Simulation}

\author{Ekaterine Dadiani$^{1}$}
\author{Tiziana Di Matteo$^{1,2}$}
\author{Nianyi Chen$^{1}$}
\author{Patrick Lachance$^{1,2}$}
\author{Yue Shen$^{3,4}$}
\author{Yu-Ching Chen$^{5}$}
\author{Rupert Croft$^{1,2}$}
\author{Yueying Ni$^{6}$}
\author{Simeon Bird$^{7}$}

\affiliation{$^{1}$ McWilliams Center for Cosmology, Department of Physics, Carnegie Mellon University, Pittsburgh, PA 15213 \\
$^{2}$ NSF AI Planning Institute for Physics of the Future, Carnegie Mellon  University, Pittsburgh, PA 15213, USA \\
${3}$ Department of Astronomy, University of Illinois at Urbana-Champaign, Urbana, IL 61801, USA \\ 
${4}$ National Center for Supercomputing Applications, University of Illinois at Urbana-Champaign, Urbana, IL 61801, USA \\
${5}$ Department of Physics and Astronomy, Johns Hopkins University, Baltimore, MD 21218, USA\\
${6}$ Harvard-Smithsonian Center for Astrophysics, 60 Garden Street, Cambridge, MA 02138, USA \\
${7}$ Department of Physics and Astronomy, University of California Riverside, 900 University Ave, Riverside, CA 92521}

\begin{abstract}
We study dual AGN host galaxy morphologies at $z=2$ using the ASTRID simulation, selecting black hole (BH) pairs with small separation ($\Delta r<30\rm{kpc}$), high mass ($M_{\text{BH,12}}>10^7M_\odot$), and luminosity ($L_{\text{bol,12}}>10^{43}\rm{erg/s}$). 
We kinematically decompose (using MORDOR) $\sim1000$ dual AGN hosts into standard components - a `disk' (thin and thick disk, pseudo-bulge) and `spheroids' (bulge and stellar spheroidal halo) and define disk-dominated galaxies by the disk-to-total $D/T\geq0.5$. 
In ASTRID, $60.9\pm2.1\%$ of dual AGN hosts (independent of separation) are disk-dominated, with the $D/T$ distribution peaking at $\sim0.7$. 
In dual-AGN hosts, the $D/T$ increases from $\sim17\% $ at $M_{\rm *}\sim 10^{9} M_{\odot}$ to $ 64\% $ for $M_{\rm *} \sim 10^{11.5} M_{\odot}$, and the pseudo-bulge is the dominant component of the disk fraction at the high mass end. 
Moreover, dual AGN hosts exhibit a higher fraction of disk/large pseudo-bulge than single-AGN hosts.
The Disk-to-Total ratio is approximately constant with BH mass or AGN luminosity.
We also create mock images of dual AGN host galaxies, employing morphological fitting software Statmorph to calculate morphological parameters and compare them with our kinematic decomposition results.
Around $83.3\pm2.4\%$ of galaxies display disk-like profiles, of which $\sim60.7\pm2.2\%$ are kinematically confirmed as disks. Se\'rsic indices and half-mass radii of dual AGN host galaxies align with observational measurements from HST at $z\sim2$. 
Around $34\%$ are identified as mergers from the $\text{Gini}-M_{20}$ relation. 
We find two dual AGN hosted by galaxies that exhibit disk-like se\'rsic index $n_{12}<1$ and $(D/T)_{12}>0.5$, which are in remarkable agreement with properties of recently discovered dual quasars in disk galaxies at $z\sim 2$.\\[0.5em]
\end{abstract}

\maketitle

\section{Introduction}

Supermassive Black Holes (SMBHs), with masses spanning from $10^6$ to $10^9 M_{\odot}$, are at the center of most massive galaxies, as inferred by both stellar dynamics and accretion \citep{1982MNRAS.200..115S, 1998AJ....115.2285M, 2013ARA&A..51..511K}. Understanding their formation and evolution is the subject of extensive research. 
Many studies confirm a tight connection between the properties of SMBHs and their host galaxy attributes such as stellar mass, galaxy bulge, star formation rate, internal velocity dispersion, etc. \citep{2004ApJ...604L..89H, 2015ApJ...813...82R, 2000ApJ...539L...9F}. This suggests a strong link between the origin and growth of SMBHs and the galaxies that reside in \citep{1998Natur.395A..14R, 2005Natur.433..604D}.

Galaxy mergers are believed to be the primary way SMBH pairs are created. After a galaxy merger, SMBHs will sink into the central regions of the newly merged galaxy due to dynamical friction. Eventually, the two SMBHs form a gravitationally bound binary pair reaching sub-parsec spacing \citep{2001ApJ...563...34M}, and continued interactions with stars and gas gradually reduce the size of their orbit and end in a merger with associated gravitational wave emission.   
The phases of a binary supermassive black hole's evolution are outlined by \cite{1980Natur.287..307B} and summarized by \cite{2003ApJ...596..860M}.  
The gravitational-wave signals produced during the final inspiral, merger, and ringdown of the binary SMBHs can be detected by projects like pulsar timing arrays \citep{2020ApJ...905L..34A, 2023arXiv230900693T} and the upcoming Laser Interferometer Space Antenna \citep{2022hgwa.bookE..17A, 2023LRR....26....2A}. 

Galaxy mergers can activate galactic nuclei (AGN) \citep{2005Natur.433..604D, 2008ApJS..175..356H}, making the binary SMBHs detectable via the electromagnetic emissions from these nuclei. In observations, dual AGNs indicate galactic-scale (tens of kiloparsecs to tens of parsecs) SMBH pairs with both black holes active. Examining dual AGNs is crucial for unraveling the origins and tracing the evolutionary path of SMBHs throughout the different epochs of the universe. However, their detection is challenging due to issues highlighted by \cite{2019NewAR..8601525D}. 
Recently, there has been significant scientific interest in identifying and researching close-separation dual AGNs \citep[e.g.][]{2023ApJ...955L..16L, 2023ApJ...945...73G, 2022ApJ...925..162C, 2023Natur.616...45C, 2022NatAs...6.1185M}. 
Most dual AGNs discovered so far are at low redshifts; there is a gap in our understanding of dual SMBHs at higher redshifts, and only a handful have been discovered to date \citep[See Figure 1 for details and references thereby][]{2022ApJ...925..162C}. 
Notably, the recent studies \citep[]{2021NatAs...5..569S, 2023ApJ...943...38S} have reported the detection of two dual AGN candidates with redshifts greater than 5, using optical spectroscopy and photometry.

Galaxies show a vast diversity of morphological characteristics that are closely related to several inherent properties, including their star formation rate and color, as well as local environment, and more \citep{2009MNRAS.398.1129G, 2010ApJ...722....1T, 2012ApJ...744...63O, 2013ApJ...762L...4L}.
Recent studies are beginning to build a census for the dual AGN population separated at galactic distances ranging from tens of kiloparsecs to tens of parsecs \citep{2023ApJ...943...38S}. 
A study by \cite{2023Natur.616...45C} unveiled the first kiloparsec-scale dual quasar, J0749, hosted by a galaxy merger at cosmic noon. 
Furthermore, companion findings from JWST NIRSpec and MIRI IFU observations reveal that the dual quasar is situated within a highly active starburst galaxy \citep{2024arXiv240308098I, 2024arXiv240304002C}. 
This finding challenges the current understanding of how supermassive black holes and their host galaxies co-evolve, as the dual quasar is hosted by two massive, compact, disk-dominated galaxies. The study proposes that some supermassive black holes may have formed before their host stellar bulges. 
Our goal is to better understand these systems from a theoretical point of view, particularly focusing on the predictions of the host galaxy morphology for dual AGN.

In this paper, we use the large-scale cosmological simulation \texttt{Astrid}, which matches the requirements for this study well \citep{Bird2022, 2022MNRAS.513..670N, 2022MNRAS.514.2220C}. 
Given its large-volume and high resolution, \texttt{Astrid} enables an in-depth examination of galaxies' internal structure and morphology.
Moreover, \texttt{Astrid} employs an improved dynamical-friction modeling, which is crucial for making accurate predictions about SMBH mergers and tracking black hole dynamics and accretion across extended timescales spanning hundreds of millions of years.

We specifically focus on studying dual AGNs at $ z \sim 2$, a starting point for forming the largest binary SMBHs (masses over $10^8 M_\odot$) following galaxy mergers. The peak of both luminous quasars and global star formation occurred around this time \citep{madau2014cosmic, 2006AJ....131.2766R}. Galaxy merger rate is much higher at cosmic noon compared to lower redshifts \citep{2019ApJ...876..110D, 2008ApJS..175..356H}. The observation of dual AGN in this epoch is also rapidly advancing thanks to the recent data from JWST NIRspec and the new observation/data processing techniques specifically targeting high-redshift dual AGN \citep[e.g.][]{Foord2020ApJ...892...29F, Perna2023}.

To study galaxy morphology in a cosmological context, an efficient and accurate method is needed to classify galaxies and break them down into their individual components. 
Several decomposition methods for galaxies have been discussed. \cite{2003ApJ...591..499A} first introduced the now widely adopted circularity parameter, while later studies, such as \cite{2012MNRAS.421.2510D}, introduced modifications and different approaches. 
More recent studies by \cite{2019ApJ...884..129D} and \cite{2022MNRAS.509.1764J} have implemented Gaussian Mixture Models and introduced a probabilistic approach for dynamic galaxy decomposition in large-volume simulations. For the purpose of this paper, we use a recent, accurate method and publicly available decomposition tool, \texttt{Mordor} (MORphological DecOmposeR) \cite{2022MNRAS.515.1524Z}, to achieve the most precise and thorough kinematic decomposition of our simulate galaxies. Additionally, it provides the capability to independently examine pseudo-bulges, an essential feature for analyzing gas-rich mergers commonly encountered at redshift 2 \citep{2020ApJ...888...65G}, a topic we will delve into subsequently.
We aim to determine what fraction of dual AGN occurs in disk versus spheroidal galaxies at $z\sim 2$. 
Additionally, we create mock images and use standard morphological fitting software Statmorph \citep{statmorph} to calculate various morphological measures as commonly used in observational studies and compare the outcomes from kinematic methods.

The paper is organized as follows. 
In Section 2, we outline the foundational methods implemented in our analysis. We briefly introduce the \texttt{Astrid} simulation and detail our selection criteria for dual AGNs at $z\sim2$.
In Section 3, we focus on the methods for decomposing galaxies. 
We review the \texttt{Mordor} tool, which we employ in this study, to examine the morphologies of simulated galaxies. We briefly summarize the standard observational galaxy classification we use on the simulated galaxy images to compare our results directly with observational findings.
Section 4 shows our results, starting with a detailed study of the various morphologies of galaxies hosting SMBH pairs. We investigate the mass fraction distributions for different kinematic structures. Following this, we study the different galaxy morphologies and examine their relationships with several properties, including the separation distance between SMBH pairs, their BH mass, and luminosity. We compare our findings with current observational data, such as the Se\'rsic indices and effective radii of the dual AGN host galaxies. We also compare the most massive and brightest dual AGNs in \texttt{Astrid} with the recent discovery of AGN pair in disk galaxy.
Finally, section 5 presents a summary of our findings along with our conclusions.

\section{Method and Approach}

\subsection{Simulations}

Few cosmological simulations can generate a significant number of AGN pairs separated by kiloparsecs at $z=2$ for various reasons, such as the necessity of a high-resolution and large cosmological volume due to the rarity of dual AGNs, which constitute only a minor percentage of the total AGN population and are even scarcer at high redshifts \citep{2011ApJ...737..101L, 2011ApJ...733..103F}. Furthermore, in many high-resolution simulations, black holes are pinned to the minimum of the gravitational potential. This leads to rapid mergers of the central massive black holes during galaxy mergers, making it challenging to observe them during the kiloparsec-separated phase.
In this study, we will use the \texttt{Astrid} simulation, which satisfies the requirements for studying high-redshift AGN pairs. First, because of its large volume, \texttt{Astrid} contains more than $10^4$ massive AGN already at redshift $z=2$ and of these $>3\%$ are in pairs \citep{Chen2022b}. Next, \texttt{Astrid} uses dynamical-friction modeling rather than repositioning (pinning the BHs to the halo minimum potential), which is very important to predict SMBH mergers accurately. It allows a considerable delay in the merger of black holes after their initial encounter, with an average time span of $\sim 200{\rm Myrs}$, and has led to one of the first studies of the evolution of  $\Delta r \sim 1 {\rm kpc}$ AGN pairs and their activation. Otherwise, two central SMBHs merge too quickly,  at  $\sim{\rm kpc}$ separations \citep[see also][]{Volonteri2022}.

The large-scale simulation \texttt{Astrid}, with box size $250 h^{-1} {\rm Mpc}$ contains $2\times5500^3$ particles.  The initial conditions are set at redshift $z=99$, and we use snapshots run to $z=2$. The simulation uses the cosmological parameter values taken from \cite{2020A&A...641A...6P} as the “base $\Lambda{\rm CDM}$” model, with matter density parameter $\Omega_m=0.32$, matter fluctuation amplitude $\sigma_8=0.81$, dark matter density  $\Omega_ch^2=0.12$, baryon density $\Omega_bh^2=0.02$ and scalar spectral index $n_s = 0.97$. The gravitational softening length is $\epsilon_g = 1.5h^{-1} {\rm ckpc}$.
Additionally, mass resolution in the initial conditions is $M_{{\rm DM}}=6.74\times10^6h^{-1}M_{\odot}$ and $M_{{\rm gas}}=1.27\times10^6h^{-1}M_{\odot}$. 

\texttt{Astrid} contains important physical processes, including models for hydrodynamics, inhomogeneous hydrogen and helium reionization, the initial velocity difference between dark matter and baryons, metal return from massive stars, star formation, supernova winds and stellar feedback, and BH seeding, dynamics and mergers. 
To solve the Euler equations, the pressure-entropy formulation of smoothed particle hydrodynamics (pSPH) is adopted as detailed by \cite{2014MNRAS.440.1865F}.
The hydrodynamics, star formation, and stellar feedback models are mostly unchanged from \cite{2016MNRAS.455.2778F}. The model for spatially inhomogeneous helium reionization follows \cite{2020JCAP...06..002B}. Supernova winds and 
stellar feedback are also included as in\citep{2010MNRAS.406..208O}.
We refer the readers to \cite{2022MNRAS.513..670N}, \cite{Bird2022}, and \cite{2022MNRAS.514.2220C} for more detailed discussion of the subgrid models and in-depth discussions regarding sub-grid models and black hole statistics.

\subsubsection{SMBH Seeding, Dynamics, and Mergers}
Here, we briefly discuss BH seeding, dynamics, and mergers. For a more detailed discussion of the SMBH model, please refer to \cite{2022MNRAS.513..670N, 2022MNRAS.514.2220C}. 

In \texttt{Astrid}, a modified SMBH seeding scheme has been adopted instead of a universal seed mass, using a power law distribution \citep{2022MNRAS.513..670N}. The FOF group finder is run periodically to find halos with specific seeding criteria according to total mass and stellar mass, with $M_{{\rm halo, FOF}}>5 \times 10^9h^{-1}M_\odot$ and $M_{*,{\rm FOF}}>2\times10^6h^{-1}M_\odot$. These specific criteria have been chosen first to have massive enough halos to contain a sufficient amount of cold, dense gas for star formation and also to have enough collisionless star particles close by for dynamical friction to be effective. For most halos, if the first criterion is satisfied, the second one is met automatically. 
This BH treatment builds on the BlueTides simulation concerning BH growth and feedback and also follows previous work by \cite{2005MNRAS.361..776S}, \cite{2005Natur.433..604D}. 

To describe the dynamics of black holes, most cosmological simulations use a repositioning method that pins BH particles to the potential minimum at each timestep and causes artificially immediate mergers of two BHs when they pass each other, even when they are not gravitationally bound. Instead of this, \texttt{Astrid} uses a sub-grid dynamical friction model, which implements a drag force that transfers energy and momentum from the SMBH to surrounding stars and, therefore, allows us to describe small-scale interactions between them \citep{2015MNRAS.451.1868T, 2017MNRAS.470.1121T, 2022ApJ...925..162C}. The total momentum is conserved by gradually transferring momentum from the SMBH to neighboring stars, and as a result, it dissipates relative velocity and slowly decays into the galaxy's center. Dynamical friction is computed using Equation $8.3$ from \cite{2008gady.book.....B}: 	
\begin{equation}
    \textbf{F}_{DF}=-16\pi^2G^2M_{BH}^2m_a\log{(\Lambda)}\frac{\textbf{V}_{BH}}{v_{BH}^3}\int_{0}^{v_{BH}}dv_av_a^2f(v_a),
\end{equation}
where $M_{\rm BH}, v_{\rm BH}$ are BH mass and velocity, $m_a, v_a$ denote masses and velocities of surrounding particles, and $f(v_a)$ is the velocity distribution function of surrounding collisionless particles, approximated by a Maxwellian distribution. The coulomb logarithm $\log{(\Lambda)}=\log{(b_{\rm max})/b_{\rm min}}$ describes the friction force range between specified  $b_{\rm max} = 20 {\rm kpc}$ and $b_{\rm min}=GM_{\rm BH}/v_{\rm BH}^2$.

\texttt{Astrid} allows us to precisely determine the dissipative velocity and acceleration at every point in time, as black holes are not restricted to potential minima while using the repositioning method, the only criterion for describing mergers is the distance between black hole pairs, which causes inaccuracy. For example, repositioning can merge BH pairs with high relative velocities, which might not be gravitationally bound and may not be eligible to merge yet or never. Using the \texttt{Astrid} simulation, we can check whether SMBH pairs are gravitationally bound or not using criteria as shown in \cite{2011ApJ...742...13B}  and \cite{2017MNRAS.470.1121T} defined by comparing kinetic energy of pairs and the work needed for BHs in pair to merge. Additionally, the simulation determines the merging distance using the gravitational softening length, below which BH dynamics is not defined because of dynamical friction:

\begin{equation}
  \left\{\begin{array}{@{}l@{}}
    \frac{1}{2}\left| \Delta \textbf{v} \right|^2 <\Delta \textbf{a}\Delta \textbf{r}\\
    \left| \Delta \textbf{r} \right| < 2\epsilon_g
  \end{array}\right.\,.
\end{equation}

Here, $\Delta\textbf{r}, \Delta\textbf{v}$, and $\Delta\textbf{a}$ correspond to the relative distance, velocity, and acceleration between the black holes in the SMBH pair, respectively. 
 
\subsection{Dual AGN Selection}
In this study, we are specifically interested in pairs of SMBHs located relatively close to each other, with a separation distance of less than $30\,{\rm kpc}$, at a redshift $2$. We restrict our analysis to pairs where both black holes have a high mass ($>10^7 M_{\odot}$) and luminosity ($>10^{43}{\rm erg/s}$) (dual AGN) to be comparable to observations. An in-depth analysis of the properties and evolution of selected dual AGN can be found in \cite{2023MNRAS.522.1895C}.
It is important to note that the luminosity criterion is applied at a single snapshot, and SMBH activity might change for close snapshots. 
However, in Fig. 3 of \cite{2023MNRAS.522.1895C}, it was shown that there is no significant dependence of dual AGN fraction on redshift from $z\sim3-2$ in \texttt{ASTRID}. Moreover, we studied the SMBH activity in nearby snapshots of a shorter timescale, from $z=1.95$ to $z=2.05$, covering 9 snapshots in total. Our analysis shows that all of the primary black holes (BH1) and $\sim98\%$ of the secondary black holes (BH2) stay active across these snapshots (mean luminosity for each BH across snapshots  $>10^{43}$erg/s), with median values of  $\sim3\times10^{44}$erg/s and $\sim7\times10^{43}$erg/s, respectively. 
Hence, the selection criterion does not affect our results, and the statistical significance remains unchanged.

We additionally divide these pairs into two groups based on their host galaxies' merging status, as determined by the subhalo catalog produced by Subfind \citep{2001NewA....6...79S}. The first group, same-galaxy pairs, includes dual AGN found within the same galaxy, while the second group, different-galaxy pairs, comprises pairs captured before their host galaxies merge. At $z=2$, about half of dual AGN pairs reside within the same galaxy, while the other half are in different galaxies.
For each pair, the black hole with the larger mass at the time of observation is referred to as the primary SMBH (or BH1), while the one with less mass is known as the secondary SMBH (or BH2). For the pairs in the same galaxy, we consider the mass and luminosity to be the sum of both BHs in the pair, not distinguishing one from another.

\begin{figure}
\centering
  \includegraphics[width=0.49\textwidth]{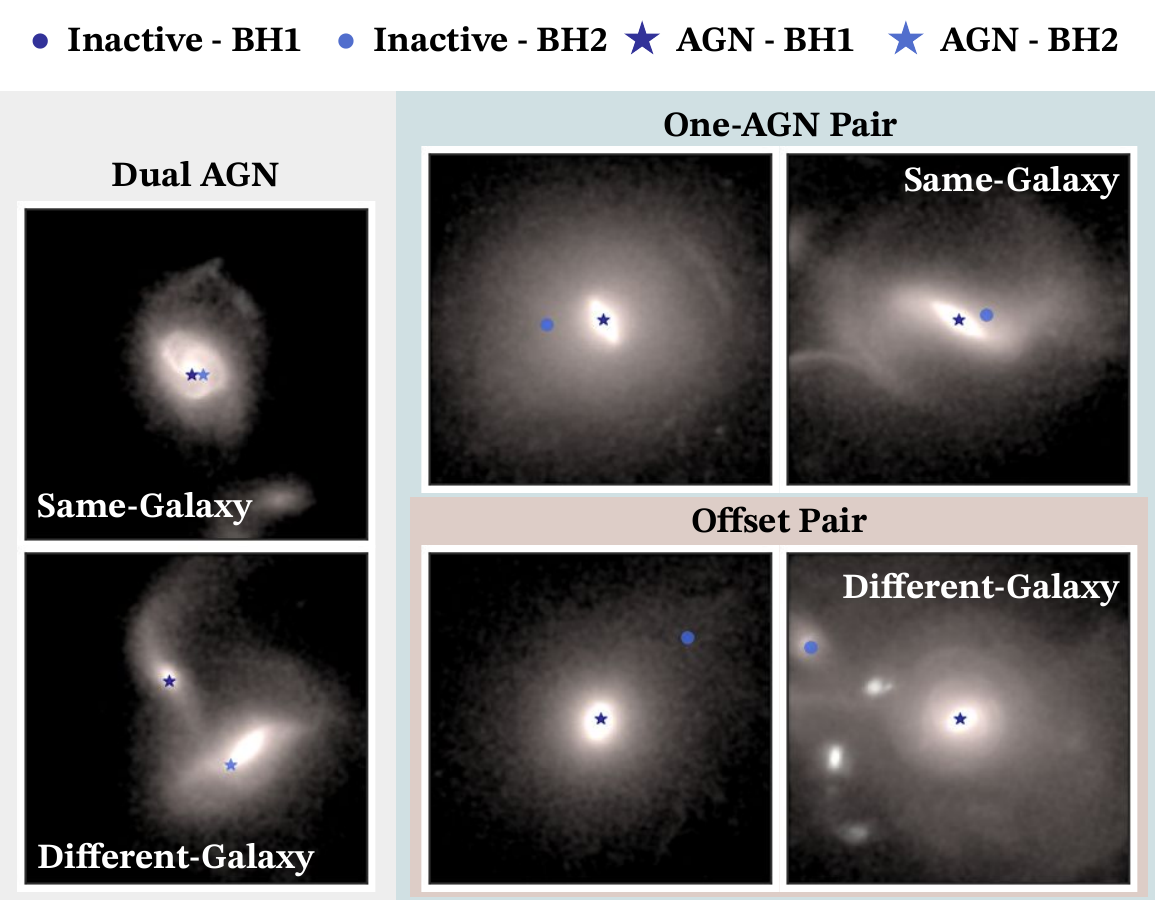}
  \caption{
 Examples of simulated dual-AGN and one-AGN pairs. The left column represents Dual AGN galaxies. The top row displays pairs with a single host galaxy, while the bottom row shows pairs with different galaxies. The right column illustrates examples of One-AGN pairs, including offset pairs displayed on the second row, which have separate galaxy hosts. Stars mark active SMBHs, while dots mark non-active ones. Darker or lighter colors indicate pairs with more or less massive SMBHs.}
  \label{dual}
\end{figure}

For completeness, in this section, we also briefly show examples of another SMBH pair population: one-AGN pairs with only one active black hole while the second SMBH in a pair remains inactive. This category includes a subset known as "offset pairs," where the black holes are located in separate host galaxies, as defined by \cite{2013ApJ...769...95B}.
Figure \ref{dual} shows examples of simulated AGN pair galaxies where supermassive black holes are represented by stars or dots referring to active and inactive black holes, respectively. 
The first column shows Dual AGN pairs, with both BHs within the same galaxy (top row) and different galaxies (bottom row). The second column features examples of One-AGN pairs, with the top row showing pairs residing in the same galaxy host for both SMBHs and the bottom row depicting its subset: Offset Pairs with different galaxy hosts. Darker and lighter stars indicate pairs with more and less massive SMBHs, respectively.

Offset AGN can be challenging to detect observationally \citep{2023ApJ...943...38S}.
In the rest of the paper, we concentrate solely on dual AGNs.
After applying the aforementioned criteria, we have identified $\sim1000$ dual AGN at redshift $z=2$. For the selected dual AGN, we study morphologies according to different static properties of BHs in pairs, such as their masses and luminosity, as well as host galaxy parameters, such as their stellar masses.

\section{Galaxy Decomposition Methods}

In this section, we describe decomposition methods to investigate the morphologies of simulated galaxies hosting SMBH pairs. 

\subsection{Kinematic Decomposition} 
 
The kinematic decomposition method initially proposed by \cite{2003ApJ...591..499A} and later refined by \cite{2009MNRAS.396..696S} segregates star particles into either disc or spheroidal components. Over time, the original formulation has undergone several modifications to increase its precision and applicability.

In the standard approach of this method, the total angular momentum of baryons in the inner halos of each galaxy is calculated first, represented as $\textbf{J}$. A reference system is then chosen, aligning $\textbf{J}$ with the z-axis. Following this, a circularity parameter is defined as the ratio of its projected angular momentum in the z-direction to the angular momentum of a circular orbit at the same radius and with the same total specific energy, denoted as $\epsilon=j_z/j_{\text{circ}}$. Here, $j_{\text{circ}} = rv_{\text{circ}} = \sqrt{GM/r}$, where $M$ represents the total mass within a radius $r$. This approach assumes that the mass distribution in the central region can be approximated as spherically symmetrical. For each star particle, the value of $\epsilon$ tends to peak at approximately 1 or 0. This indicates whether the particle is part of a disk component characterized by rotational motion or a spheroidal component dominated by velocity dispersion. 
  
This definition of circularity parameter $\epsilon$ has been used widely in literature and has been improved using additional constraints to be compatible with observational results. Though computationally efficient, this method can introduce inconsistencies dependent on the galaxy's structure, especially for very disturbed galaxies. In this study, we adopt a more precise method, a publicly accessible decomposition tool, \texttt{Mordor} (MORphological DecOmposeR), presented in \cite{2022MNRAS.515.1524Z}, to accomplish a more accurate and comprehensive kinematic decomposition. This is based on two key studies: the above-mentioned work of  \cite{2003ApJ...591..499A} who first introduced the concept of the circularity parameter, and \cite{2012MNRAS.421.2510D} who emphasized the role of stars' binding energy. 

\subsubsection{Mordor}
The \texttt{Mordor} algorithm is available at \url{https://github.com/thanatom/Mordor}. We provide a brief overview of this method here; for an in-depth understanding of the \texttt{Mordor} technique, please refer to  \cite{2022MNRAS.515.1524Z}.

The kinematic decomposition of galaxies relies fundamentally on two key quantities associated with the star particles used for sampling: the total specific energy ($E$) and the circularity parameter ($\epsilon$). The total specific energy is straightforwardly determined using the equation \( E = \frac{1}{2} |v|^2 + \epsilon_g \), where \( v \) denotes the particle velocity and \( \epsilon_g \) refers to the gravitational potential. Meanwhile, the calculation of circularity requires a more meticulous approach. It is computed as \( \epsilon = \frac{j_z}{j_{\text{circ}}(E)} \), where \( E \) is the orbital energy and \( j_{\text{circ}}(E) \) is calculated under the presumption of axial symmetry in the equatorial plane of the potential. It also necessitates the evaluation of the gravitational potential to estimate the circularity for all the chosen stellar particles,  a vital aspect of this method.

The employed decomposition algorithm identifies five distinct structural components within galaxies: a thin disc, a thick disc, a pseudo-bulge, a central spheroid (also known as the bulge), and a less tightly bound stellar spheroidal halo. (This five-part framework is comparable to that in \cite{2021ApJ...919..135D}. However, it does not rely on a Gaussian fitting machine learning algorithm.)

The decomposition process involves a series of steps: extracting particle data sets, calculating gravitational potentials, recentering galaxies, estimating the circular angular momentum and total energy profiles in the galactic plane, eliminating potential outliers, binning the particle energy distributions, identifying the minima (termed $E_{cut}$), and finally, segregating stars into more ($E\leq E_{cut}$) or less ($E> E_{cut}$) bound entities. The \texttt{Mordor} algorithm then analyzes the most bound components,  categorizing them as either bulge, thin disc, or pseudo-bulge based on their circularity. Particles with $\epsilon<0$ are assigned to the bulge, mirroring this distribution for positive circularity. Among the remaining particles, those with $\epsilon>0.7$ are assigned to the thin disc, while the rest are associated with the pseudo-bulge. A similar approach is used for less bound particles, categorizing them into a stellar halo, thin disc, or thick disc. 
It is also important to note the potential influence of the bar structures. \cite{2022MNRAS.515.1524Z} provides an in-depth bar identification analysis and looks into details of their properties. However, we do not conduct a distinct bar identification in this study. Thus, stars in bars might be attributed to either the pseudo-bulge or the spheroidal component, given their distinct angular momentum relative to the disc.

When compared to other algorithms described in the literature, \texttt{Mordor} shows good agreement and greater flexibility, accurately decomposing systems.

After the decomposition of galaxies into five distinct kinematic components, we calculate the disc-to-total ratio as \( D/T = \frac{M_{\text{Disk}}}{M_{*}} \). 
The concept of a "disc galaxy" traditionally describes systems where the majority of stellar orbits exhibit aligned angular momentum. However, the system's physical characteristics can significantly influence the actual classification of a disc galaxy. Given the morphological diversity observed in real galaxies, setting a fixed threshold for this classification is challenging. Moreover, it should also be noted that distinguishing between pseudo-bulges and classical bulges using observational techniques is challenging. While several methods, including stellar kinematic analyses and photometric profiling, have been employed, they have not yet provided unambiguous identification. This challenge has been discussed in detail in studies such as \cite{2022MNRAS.515.1524Z}. The formation of pseudo-bulges is believed to result from secular evolution driven by non-axisymmetric features in a galactic disk, including elements like spiral arms and bar-like structures  \citep{2005MNRAS.358.1477A, 2012MNRAS.422.1902I}. \cite{2004ARA&A..42..603K} theorized that pseudo-bulges preserve characteristics of their original disk formation. This is supported by observations that reveal disk-like properties in pseudo-bulges, such as an exponential density profile, a flatter shape than classical bulges, and the prevalent presence of central features like bars, spirals, and rings.

For the purpose of this study, we define the disk mass as the combined mass of all rotational components – namely, the thin disc, the thick disc, and the pseudo-bulge: \( M_{\text{Disk}} = M_{\text{Thin Disk}} +M_{\text{Thick Disk}} +M_{\text{Pseudo Bulge}} \),  and bulge as the combined mass of bulge mass (sometimes referred as spheroidal mass), and stellar halo: \( M_{\text{Bulge}} = M_{\text{Bulge}} +M_{\text{Halo}}\).
The threshold for the disk-to-total ratio to determine disc-dominated galaxies differs across various studies. 
We categorize any galaxy with a \( D/T =M_{\text{Disk}}/M_{*} \geq 0.5 \) as a disc galaxy and \( D/T < 0.5 \) as a bulge galaxy. 

\begin{figure*}
\centering
  \includegraphics[width=0.99\textwidth]{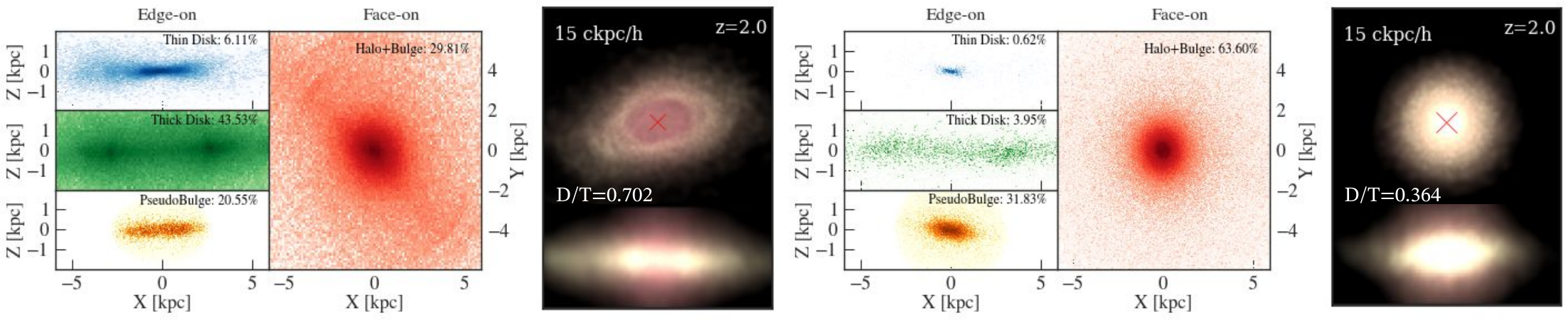}
  \caption{  
  Classification of the galaxy structures hosting a single SMBH from dual AGN pairs based on \texttt{Mordor} decomposition for disk (left panel) and spheroidal (right panel) galaxies. In each panel, the first columns show edge-on projections of the disk components (thin disk, thick disk, and pseudo-bulge), along with their relative contributions in the upper right corner. The second column displays a face-on projection of the combined spheroidal components (bulge and halo). Additionally, the far-right column shows the smoothed stellar density for the entire stellar distribution. Color-coded by stellar age (warmer colors correspond to older stars)}
  \label{Mordor}
\end{figure*}

Figure \ref{Mordor} shows the kinematic decomposition of two example galaxies using the \texttt{Mordor} tool. Here, we show two contrasting examples: the first figure illustrates a disk galaxy with a disk-to-total ratio of \(0.702\). In contrast, the second displays a spheroidal galaxy with a D/T ratio of \(0.364\).

Each component of the five-part decomposition is shown separately to clarify their individual contributions. In each panel, the left columns present edge-on projections of components contributing to rotational movement, which we categorize as disk stars: the thin disk, thick disk, and pseudo-bulge, along with their respective mass fractions. On the other hand, the right columns show the face-on projection of the combined spheroidal components, namely the central bulge and the surrounding halo. From the first figure, it is apparent that the disk components have a significant presence, with contributions of around \(6\%\), \(40\%\), and \(20\%\) from the thick disk, thin disk, and pseudo-bulge, respectively. In contrast, the second figure highlights the dominance of spheroidal components, constituting \(\sim63.6\%\) for the bulge and halo components together.

Additionally, the figures include maps of the stellar surface density of both face-on and edge-on projections, encompassing all stellar components. These maps clearly indicate that the first galaxy predominantly exhibits disk characteristics, whereas the second galaxy is significantly dominated by its bulge.

\subsection{Mock observations and morphology fitting}
\label{sec: mock observations}

In addition to kinematic decomposition and stellar density maps, we made simple mock observations of the galaxies in our sample in order to make morphological fits and compare those with observations.

To create these mock observations, we identify all of the star particles associated with each galaxy within \texttt{Astrid} and use their characteristics to calculate a spectral energy density (SED) for each star particle. We use the Binary Population and Spectral Population Synthesis model \citep[BPASS version 2.2.1;][]{Stanway2018}, which includes the age, mass, and metalicity of each star particle to calculate its SED. These SEDs are each convolved with a mock HST WFC3 f160w filter to produce the total luminosity detected through that filter for each star particle. This filter was chosen in order to best match our mock images to the observations in \citet{2023Natur.616...45C}. We use the \texttt{SynthObs} code package to carry out the SED assignment and filter application portion of this process \citep{Wilkins17, FLARESII}.

Next, we create mock images based on these SEDs using the cosmological imaging software \texttt{Gaepsi2}. In addition to the positions and luminosities of each star particle, we provide the smoothing length of each particle. This is the distance encompassing the nearest $60$ star particles, which is used to assign an SPH kernel to each particle to distribute its light based on the density of its environment. We chose a pixel size of 0.060" in order to emulate the drizzled images used in \citet{2023Natur.616...45C}.

We passed these mock images to the morphological fitting software \texttt{Statmorph} \citep{statmorph} in order to calculate a variety of morphological measures. These include fitting a 2D-se\'rsic profile and calculating non-parametric statistics like the Concentration, Asymmetry, and Smoothness (CAS) statistics \citep{Bershady_00, Conselice_03}, the Gini coefficient and $M_{20}$ values \citep{2003ApJ...588..218A, 2004AJ....128..163L}.
\section{Results}
\label{sec:results}
In this section, we examine the morphological characteristics of selected SMBH pairs at $z = 2$. We analyze the distribution of mass within various kinematic structures of these galaxies decomposed using the \texttt{Mordor} algorithm. We investigate the connection between the separation of the dual AGN and their host galaxy's morphology, as well as the impact of black hole mass and luminosity on these structures. We compare our sample with observational data, particularly emphasizing the most massive, high-luminosity dual AGNs.

\subsection{Host galaxy morphology of Dual AGN}

\begin{figure*}
\centering
  \includegraphics[width=0.9\textwidth]{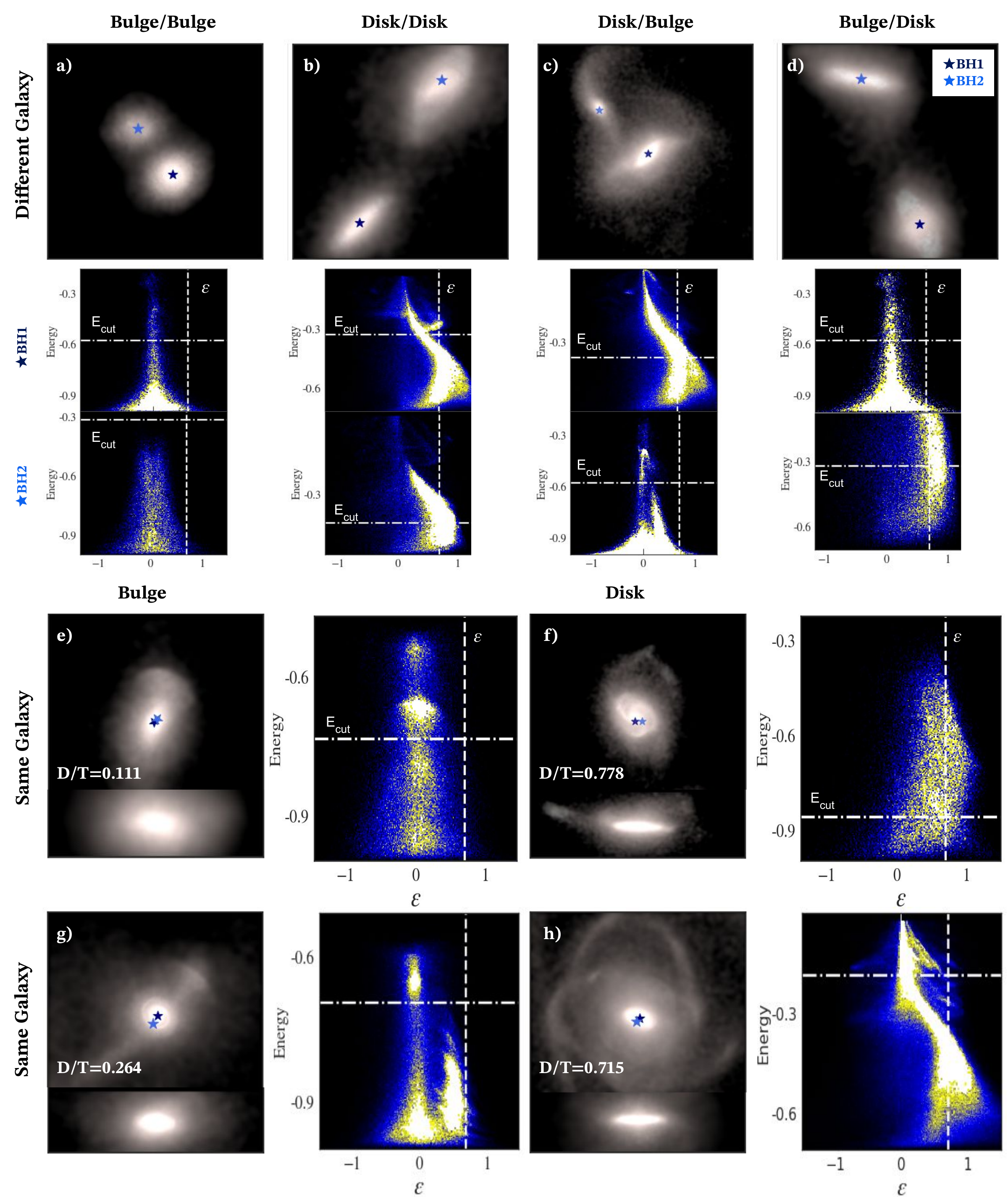}
  \caption{
  Stellar density maps and energy vs. $\epsilon$ plots of simulated dual AGN galaxies with varying morphologies. 
  The first row (panels a-d) shows pairs with different host galaxies: (a) both SMBHs in bulge-dominated galaxies, (b) both in disk-dominated galaxies, (c) primary SMBH in a disk and secondary in a spheroid, (d) primary SMBH in a spheroid and secondary in a disk. 
  Below each map are corresponding energy vs. $\epsilon$ plots for all star particles in both the primary and secondary galaxies, simulated by \texttt{Mordor}. The vertical dashed line shows threshold $\epsilon=0.7$ and the dot-dashed horizontal line corresponds to $E_{\text{cut}}$ value.
  The subsequent rows depict pairs within a single host galaxy: (e) and (g) in disk-dominated galaxies, (f) and (h) in bulge-dominated galaxies, showcasing both face-on and edge-on projections followed by their respective energy vs. $\epsilon$ plots.}
  \label{epsilon}
\end{figure*}

Galaxy mergers can significantly alter the morphology of galaxies, making it challenging to accurately decompose the galaxies during this process. In \cite{2022MNRAS.515.1524Z} \texttt{MORDOR} has been applied to all of the galaxies in TNG50, and the paper comments on the reliability of this method to high redshift merging systems. 
Our findings broadly align with theirs, particularly noting that post-z=2, the most massive galaxies exhibit significant morphological changes due to major mergers, with the median mass fraction of pseudo-bulge component being most prominent at $z\approx2$, showing higher contribution compared to the smaller mass galaxies. While, the bulge component in these massive galaxies remains relatively low at $z=2$ and begins to increase only at higher redshifts.
Here, we apply \texttt{Mordor} analysis to the hosts of dual AGN host galaxies, which reside in a subsample of relative massive galaxies involved in a galaxy merger. Given our relatively large sample, we show a series of example galaxies at different interaction stages.

Figure \ref{epsilon} illustrates the application of the \texttt{Mordor} decomposition method to dual AGN merging systems, showing a range of simulated examples of galaxies with varied morphologies.

The top row shows stellar density maps of simulated galaxies hosting dual AGN, where supermassive black holes in pairs are in different galaxies at the time of observation. The dark blue and light blue stars indicate primary and secondary BHs in a pair, respectively. Below each of the density map we show the energy vs. circularity of the primary and secondary SMBH host galaxies. The vertical dashed line shows threshold $\epsilon=0.7$ and the dot-dashed horizontal line corresponds to $E_{\text{cut}}$ value, which determines whether particles are considered more or less bound. These SMBHs are in the early stages of a galaxy merger.
As a result, they are not significantly disturbed by each other. 

Moving from left to right, there are four distinct situations: 
Panel (a) features a pair with both SMBHs in galaxies dominated by a central bulge. The symmetry around $\epsilon = 0$ in the $E$ vs $\epsilon$ plots indicates classical non-rotating bulges and minimal disturbance.
Panel (b) shows both SMBHs in galaxies with prominent disk structures. Particles with circularity parameters $\epsilon > 0.7$ indicate thin disks, while particles with  $\epsilon < 0.7$ diverging from symmetry around $\epsilon = 0$ suggest disk components such as pseudo-bulges ($E < E_{cut}$) and thick disks ($E > E_{cut}$), with minimal bulge or halo presence.
Panel (c) presents a pair where the more massive SMBH is in a disk galaxy and the less massive one in a spheroidal galaxy. In the disk galaxy (BH1), $\epsilon > 0.7$ signals a thin disk, while $\epsilon < 0.7$ splits into symmetrically distributed less bound stellar halo and non-symmetrically distributed pseudo-bulge and thick disk components. Conversely, the spheroidal galaxy (BH2) mostly shows symmetric $\epsilon$ distributions, with bound particles indicating bulge components and unbound ones suggesting a stellar halo. Non-symmetric $\epsilon < 0.7$ bound particles also denote pseudo-bulge components, indicating dynamic disturbances of galaxies as also reflected in the stellar density maps.
Panel (d) depicts the opposite scenario, where the spheroidal galaxy hosting BH1 shows a symmetric $\epsilon$ distribution around zero, with bound particles representing bulge components and fewer unbound ones suggesting halo components. There are no disk components. In contrast, the disk galaxy hosting BH2 lacks symmetry around $\epsilon = 0$, displaying a thin disk ($\epsilon > 0.7$), pseudo-bulge ($\epsilon < 0.7$ and $E < E_{cut}$), and thick disk ($\epsilon < 0.7$ and $E > E_{cut}$), indicating minimal disturbance between pair.

In the last two rows, we show examples where SMBH pairs are found within a single galaxy, typically at the onset of merging, exhibiting significant disturbances as evidenced by noticeable tails.
Panels (e) and (g) depict a disk-dominated mutual galaxy with the upper part showing a face-on projection and the lower part an edge-on projection. In panel (e), the $\epsilon$ distribution peaks near $0$, signaling rotational dominance with a $D/T$ of approximately $0.111$. Panel (g) also reveals a strong, symmetric $\epsilon$ distribution around $0$, indicating central bulge ($E<E_{cut}$) and stellar halo ($E>E_{cut}$) components, along with a notable pseudo-bulge component ($\epsilon<0.7$), yet maintaining a rotationally dominated structure with a $D/T$ of about $0.264$.
Conversely, panels (f) and (h) display a bulge-dominated mutual galaxy with corresponding face-on and edge-on projections. In panel (f), the $\epsilon$ distribution peaks close to $1$, indicating a disk-dominated structure with a $D/T$ of roughly $0.778$. Panel (h) shows a pronounced $\epsilon$ peak at $1$, highlighting disk components like a thin disk ($\epsilon > 0.7$) and a pseudo-bulge ($\epsilon < 0.7$ and $E < E_{cut}$), with symmetry around $\epsilon = 0$ suggesting the presence of both bulge and halo components. Despite these, the galaxy remains primarily disk-dominated ($D/T = 0.715$), though the significant bulge and halo elements alongside dominant disk structures suggest a more disturbed system than typically seen in disk-dominated galaxies.

The decomposition algorithm might face some challenges and might be slightly less precise with galaxies that have recently merged and are very morphologically disturbed, such as galaxies shown on the (g) and (h) panels. Despite potential challenges, we see that the epsilon parameter effectively captures vital characteristics of recently merged galaxies as shown in the provided examples, affirming the \texttt{Mordor} algorithm's ability to yield reliable insights even with highly disturbed galaxies, suggesting its suitability for the analysis performed in this study.

\subsection{Mass fractions of kinematic structures}
\begin{figure}
\centering
  \includegraphics[width=0.49\textwidth]{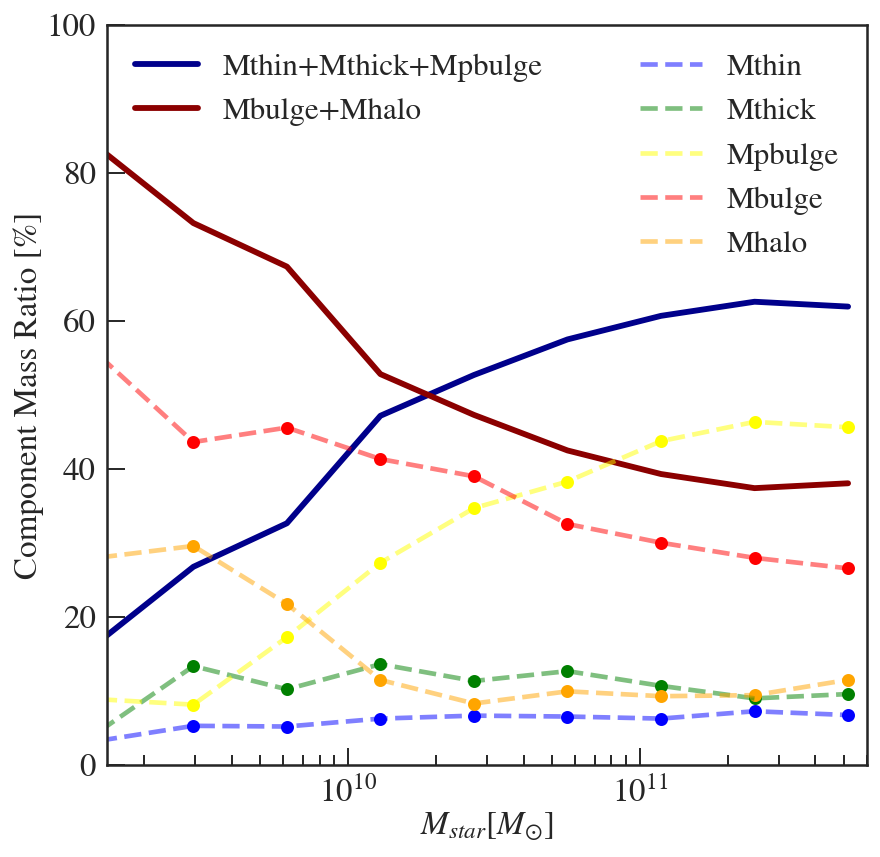}
  \caption{The dashed lines show the mass fraction of the five kinematic structures - Thin Disk (\textit{blue}), Thick Disk (\textit{green}), Pseudo Bulge (\textit{yellow}), Bulge (\textit{red}), and Halo (\textit{orange})- identified by \texttt{Mordor}. 
  For each component within a stellar mass bin, we calculate the ratio of its cumulative mass to the bin's total stellar mass. Dashed curves represent these ratios, with component-specific colors shown in the plot's label. Solid curves depict mass fractions for the Disk (blue) and Spheroidal (dark red) components, consistent with prior definitions.}
  \label{mass_comp}
\end{figure}

In this subsection, we further examine the mass fractions of the various kinematic structures as decomposed by the \texttt{Mordor} algorithm, illustrated in Figure \ref{Mordor}. Figure \ref{mass_comp} displays the mass fractions associated with each of the five structural components: the thin disk, thick disk, pseudo-bulge, bulge, and halo for galaxies hosting dual AGN with stellar masses ranging from $10^{9}$ to $10^{11.5}M_\odot$ (galaxies exceeding this mass constitute less than $1\%$ of the sample). For each stellar mass bin, we calculate the ratio of the cumulative mass of a given component in a bin to the total stellar mass of that bin, shown by dashed lines.
 
We observe a decrease in the spheroid mass fraction (a composite of bulge and halo, illustrated by the dark red solid line) from around $83\%$ to $36\%$ as the galaxy stellar mass increases from $10^9M_{\odot}$ to $10^{11.5}M_\odot$. 
The significant contribution to this trend comes from the bulge fraction, which, as indicated by the red dashed line, decreases sharply from around $0.50$ to approximately $0.27$ as the galaxy's stellar mass increases. The mass fraction of halos, represented by the orange dashed curve, shows a decrease from $\sim33\%$ at the lower end of the stellar mass spectrum, remaining relatively constant at around $8\%$ beyond $\sim10^{10}M_\odot$ stellar mass.

Based on the definition mentioned above, the disk mass comprises three components: the thin disk, the thick disk, and the pseudo-bulge. Contrasting this with the spheroid mass, there is a discernible increase in the total disk mass fraction. This begins at approximately $17\%$ and escalates to nearly $64\%$ across our galaxy's stellar mass range, as illustrated by the blue curve.
Elaborating on individual components,  the mass fraction of the thin disk (indicated by the blue dashed curve) remains relatively small (less than $10\%$) independent of the galaxy's stellar masses and contributes minimally to the total disk mass. 
The thick disk, represented by the green dashed line, undergoes a slight increase in its mass fraction with the growth in galaxy stellar mass yet remains below $20\%$.
The mass fraction of the pseudo-bulge, denoted by the yellow dashed curve, experiences a noticeable increase ranging from $8$ to $47$ percent within our designated galaxy stellar mass range. This makes it the primary component contributing to the disk mass. 

The increasing fraction of the pseudo-bulge in galaxies at the highest mass aligns with expectations, considering that our sample encompasses gas-rich mergers typically observed at redshift $z=2$  \citep{2020ApJ...888...65G}. 
\cite{2011ApJ...733L..47F} found that pseudo-bulges are notably widespread, existing in about $75\%$ of bulge-disk galaxies. Their kinematic properties resemble disks, setting them apart from classical bulges \citep{2012ApJ...754...67F}.
Pseudo-bulges are distinguished by their flatter, less dense structures and the common occurrence of complex central formations such as bars, disks, and spirals, as detailed by \cite{2004ARA&A..42..603K}.
 Additionally, they generally exhibit lower Se\'rsic indices, a characteristic of disk-like formations \citep{2008AJ....136..773F, 2010ApJ...716..942F, 2009MNRAS.398.1129G}.

Furthermore, in this study, we do not differentiate disk galaxies into barred and unbarred subsets. As a result, stars found within bars may predominantly be classified as being in the pseudo-bulge. Since bars originate from the stellar disk, it is appropriate to consider them part of the disk structure. As per  \cite{2022MNRAS.515.1524Z}, the fraction of galaxies with bars at $z=0$ rises with mass and peaks at a stellar mass of $10^{11}M_\odot$.  In the mass range of $3\times10^9-3\times 10^{11} M_\odot$, strong bars are common and account for about $50\%$ of disc galaxies at around $\sim4\times10^{10}M\odot$. This stellar mass range matches where we see the pseudo-bulge abundance. 
It is also worth noting that for galaxies with stellar masses less than $3 \times 10^9 M_\odot$, the mass fraction of the thick disk surpasses that of the pseudo-bulge.

\begin{figure}
\centering
  \includegraphics[width=0.49\textwidth]{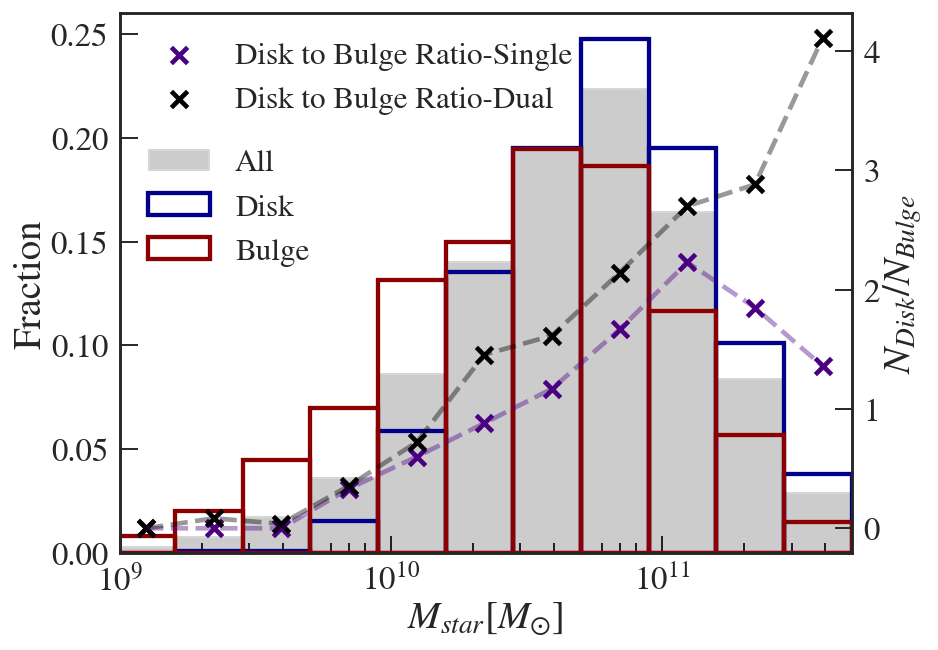}
  \caption{ 
  Stellar mass distribution for different morphology galaxies.
  The distribution of the host galaxies for the entire dual AGN sample is displayed in the grey-shaded histogram. The blue and red histograms represent the distribution of disk-dominated and bulge-dominated galaxies among them, respectively. The black crosses indicate the ratio of disk galaxy counts to spheroidal galaxy counts for each stellar mass bin for dual AGN. Similarly, the purple crosses show the same ratio for host galaxies of a sample of approximately $1000$ single AGNs at $z=2$. The sample was selected randomly to ensure a nearly uniform distribution of SMBH masses.
  }
  \label{spheroid}
\end{figure}

In Figure \ref{spheroid}, we show the stellar mass distribution for galaxies with different morphologies. The grey histogram shows the overall distribution of stellar masses for galaxies hosting selected dual AGNs, showing a notable peak at $\sim10^{10.85} M_\odot$. Among them, we separate disk-dominated and bulge-dominated galaxies, using $D/T=0.5$ as a threshold criterion, and show their stellar mass distributions separately with blue and red histograms. The number of disk galaxies increases with increasing stellar mass, reaching its peak at $\sim 10^{10.85}M_\odot$, and then decreases for higher stellar mass galaxies, following the trend of the overall count of massive galaxies. 
In contrast, spheroidal galaxies show a more gradual increase and peak at a slightly lower stellar mass bin compared to the disk distribution at $10^{10.59}M_\odot$, followed by a sharp drop. 

Moreover, we calculate the number ratio of disk galaxies to spheroidal galaxies within each stellar mass bin, marked by black crosses. Spheroidal galaxies outnumber disk galaxies up to a stellar mass of $2\times 10^{10}M_\odot$, as also indicated in figure \ref{mass_comp}. However, as we move towards the higher stellar masses, the ratio of disk galaxies to spheroidal galaxies increases. At $10^{11.5}M_\odot$, the ratio reaches a peak of approximately $3.8\pm1.5$, indicating that the dual AGN host galaxies with higher stellar masses are dominated by disk galaxies. These findings align with previous discussions on mass ratios.

We compare the mass distribution of host galaxies for dual AGN with a group of $\sim1000$ single AGN at $z=2$. The single AGNs are selected randomly such that their SMBH masses are evenly distributed in the range of $10^7 - 10^{10} M_\odot$, and they all have a high luminosity of over $10^{43}\rm{erg/s}$. In the figure, the purple crosses represent the number ratio disk to spheroidal galaxies among single AGN hosts across different stellar mass bins. Initially, the ratio increases similarly to dual AGNs, reaching a peak value of around $2.3$ for the stellar masses near $10^{11}M_\odot$. However, in contrast to dual AGNs, the ratio starts to decrease in populations of higher stellar mass galaxies.

Therefore, in massive systems, dual AGN host galaxies show a more pronounced indication of disk-like characteristics and pseudo-bulge formation than single AGN populations. This is expected, as the presence of a pseudo-bulge often signals a recent merger, and on average, dual AGNs are typically closer to such a merger event compared to single AGNs.

For the smaller mass galaxies, distinguishing dual AGN systems may become trickier; what might appear as a single AGN could, in fact, be an offset AGN, with a second AGN that is not luminous enough to meet the criteria for dual AGN classification. As a result, in these lower-mass galaxies, single and dual AGN systems can seem similar. As a result, for low-mass galaxies (up to $10^{11}M_\odot$), the disk-to-bulge ratio trend in single AGN host galaxies follows that of the dual AGN population, showing an increase with increasing galaxy stellar mass, albeit at a slightly lower rate everywhere.

\subsection{Disk-to-total ratio distribution}

\begin{figure}
\centering
  \includegraphics[width=0.49\textwidth]{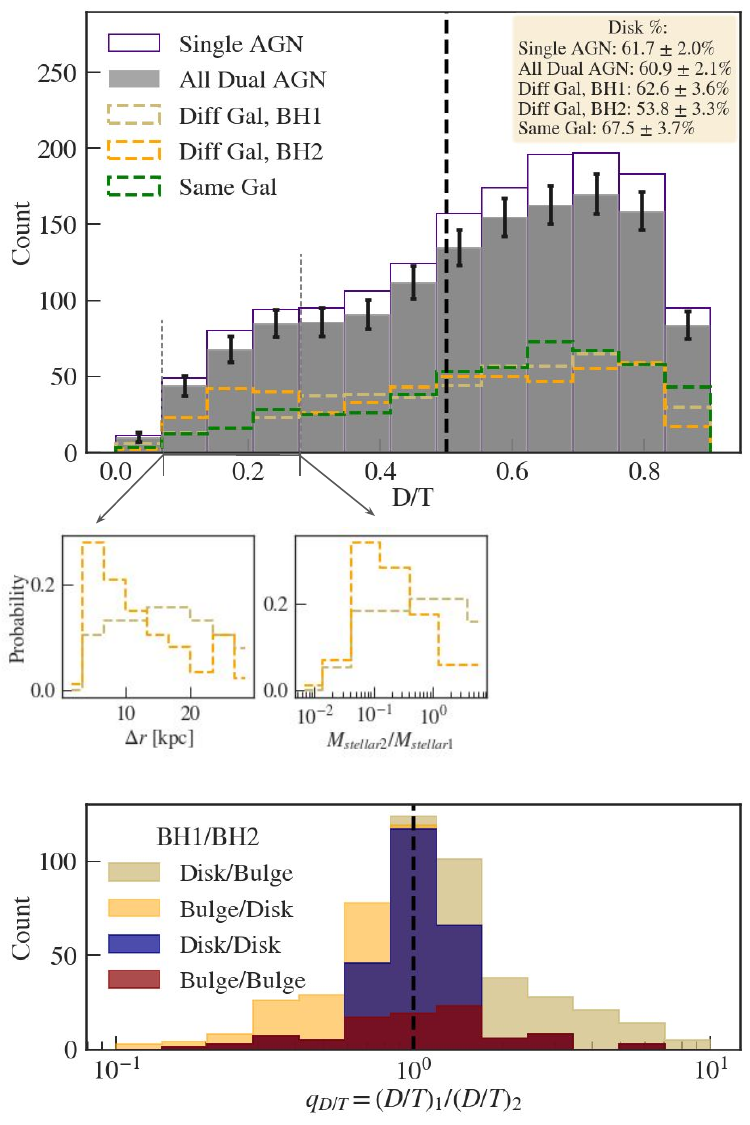}
  \caption{
  \textbf{Top:} The distribution of the disk-to-total ratio. The grey-shaded histogram represents the distribution for dual AGN. The green histogram illustrates pairs with both SMBHs in a single galaxy. Dual AGN, residing in separate host galaxies, are illustrated with dashed orange lines for the more massive SMBHs and wheat-colored lines for the less massive SMBHs. The purple histogram shows the distribution of D/T for the selected sample of single AGNs.
  \\
  \textbf{Middle:} The distribution of separation distances (first panel) and stellar mass ratios (second panel) for pairs where the primary black hole (BH1) host galaxy falls within the D/T range of $[0.1, 0.3]$ (orange), compared to those where the secondary black hole (BH2) host galaxy falls within the same range (wheat-colored).
  \\
  \textbf{Bottom:} Distribution of the ratio of D/T between the less massive and more massive SMBHs in different-galaxy dual pairs. 
  Red and blue-filled areas depict histograms for pairs where both black holes are located in the bulge-dominated and disk-dominated galaxies, respectively. Pairs where BH1 is in a disk-dominated and BH2 is in a bulge-dominated galaxy are represented by the wheat-colored area, while the reverse scenario is highlighted in orange.}
  \label{q_DT}
\end{figure}

In figure \ref{q_DT}, in the upper panel, we plot the distribution of the disk-to-total  (D/T) ratio values for dual AGN host galaxies, shown by the grey shaded area. The vertical black dashed line indicates the threshold between disk-dominated and bulge-dominated galaxies. The majority of Dual AGN, $\sim61\%$, are found in disk-dominated galaxies, with the distribution reaching a peak at $\text{D/T}\sim0.7$.

We further categorize these pairs based on whether, at the time of observation, the two SMBHs in a pair are within a single galaxy, termed as "same-galaxy pairs", or in separate galaxies, referred to as "different-galaxy pairs". The green histogram shows the D/T distribution for the same-galaxy pairs, which accounts for $\sim50\%$ of all dual AGN. This distribution follows a trend similar to the grey histogram, with $\sim68\%$ classified as disk galaxies. 

We further differentiate the different-galaxy pairs based on the mass of the black hole they host. The orange histogram represents galaxies hosting the more massive black holes, referred to as primary BHs, while the wheat-colored histogram represents those with the less massive ones, termed secondary BHs.  
The majority of galaxies hosting the primary black hole are disk galaxies, constituting $\sim63\%$ of the sample. In contrast, $D/T$ distribution for galaxies hosting the secondary BH does not show a distinct peak, with roughly $\sim54\%$ classified as disk galaxies. 
This difference is particularly notable in the $0.1<D/T<0.3$ range, where BH2 host galaxies tend to be more spheroidal than those hosting BH1. 
Further comparison of these pairs is shown in the middle panel of Fig. \ref{q_DT}. The analysis shows that pairs with secondary BH host galaxies within this D/T range (orange) tend to have closer separation distances and lower stellar mass ratios ($M_{\text{stellar2}}/M_{\text{stellar1}}$) compared to those pairs with primary BH host galaxies in the same range (wheat-color). This suggests that the higher occurrence of spheroidal galaxies hosting BH2 could be influenced by tidal disruptions from their primary counterparts galaxy, which could affect their stellar structure in the early merger stages.

Additionally, the purple histogram displays how the disk-to-total ratio is distributed among the host galaxies of the single AGN sample mentioned earlier. This distribution is consistent with the D/T of dual AGN host galaxies, as confirmed by the Kolmogorov-Smirnov (K-S) test, which shows a K-S statistic of $D=0.0094$ and a $p$-value of $0.99$. 
Approximately $62\%$ of single AGN host galaxies are disk-dominated.

The lower panel shows the distribution of D/T value ratios between the more and less massive SMBHs in a pair hosted by different galaxies: $q_{D/T}={(D/T)_1}/{(D/T)_2}$. This histogram peaks around $1$, suggesting that a majority of these pairs have host galaxies with similar morphologies. 

We further classify them based on the morphology of galaxies hosting the BHs in a pair.
Pairs, where both black holes are in bulge-dominated galaxies, are shown by the red-shaded area. This scenario is the least common, accounting for roughly $19\%$ of these pairs. 
The relatively low rate of 'bulge/bulge' systems can be influenced by the dominance of disk galaxies among AGN hosts at this redshift. As illustrated in the top panel of Figure \ref{q_DT}, around $62\%$ of the single AGN hosts we identified have disk-like morphologies. Furthermore, observational data from the HST WFC3/IR surveys (\cite{2011ApJ...727L..31S}) show that approximately $80\%$ of AGN host galaxies at $z\sim2$ have disk-like structures, suggesting that the AGN population at this redshift is predominantly gas-rich and disk-dominated.
However, such a low rate of 'bulge/bulge' systems is mainly due to merging systems' characteristics, where the most prominent feature is often the pseudo-bulge, as previously demonstrated in Figure \ref{mass_comp}. Thus, the scarcity of such mergers is due to the intrinsic properties of dual AGN host galaxies at this redshift.

When both black holes in a pair are in disk-dominated galaxies, they are depicted by the blue-shaded area. This is the most frequent scenario, comprising about $35\%$ of the pairs.
There is also a substantial number of pairs, about $48\%$, where galaxies of different morphologies host the black holes in a pair.
Cases where the primary black hole is in a disk galaxy while the secondary is in a spheroidal galaxy. This configuration makes up approximately half of this subset. The inverse, where the primary black hole is in a spheroidal galaxy and the secondary is in a disk galaxy, is denoted by the light wheat-shaded area, making up the other half of this subset.

\subsection{Properties of Disk and Spheroidal galaxies}
\subsubsection{Morphology as a function of SMBH pair separation distance}

\begin{figure}
\centering
  \includegraphics[width=0.49\textwidth]{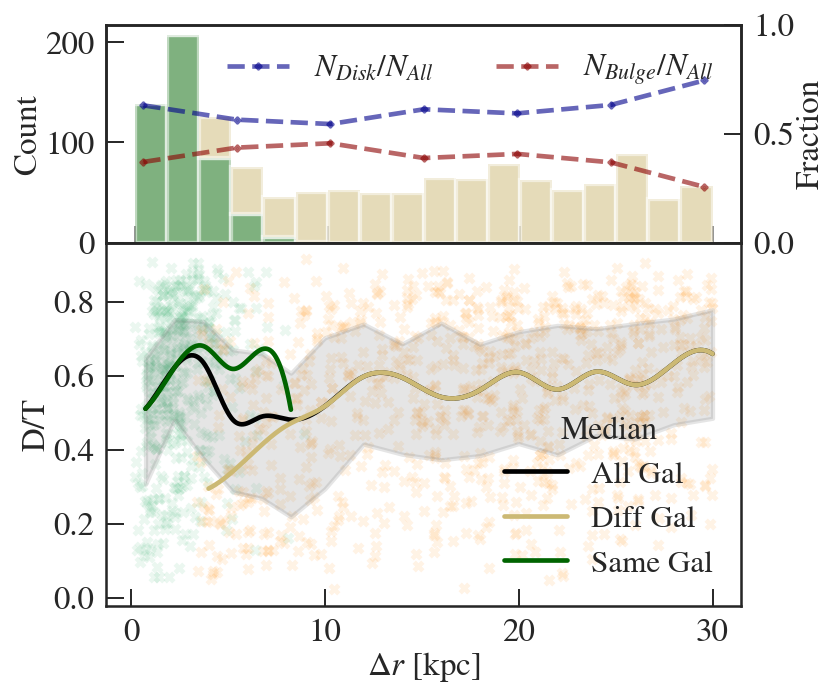}
  \caption{ 
  The disc-to-total (D/T) ratio vs the separation distance ($\Delta r$) for dual AGN. Green/orange crosses indicate pairs with BHs in the same/different galaxies. A black curve shows the median D/T values for each $\Delta r$ bin for all galaxies and green and orange curves for pairs with the same and different host galaxies, respectively.
  In the top panel, the distribution of separation distances is displayed. The blue and red dashed lines show the ratio of disk-to-total and bulge-to-total galaxy counts, respectively, within each $\Delta r$ bin. }
  \label{dr}
\end{figure}

In Figure \ref{dr}, we show the relation between the separation distance of black holes in dual AGN pairs, $\Delta r$, and the morphologies of their host galaxies.

The bottom panel shows the scattered relationship between $D/T$ and $\Delta r$. Pairs within the same/different galaxies are marked with green/orange crosses, respectively. We see that the data points are very dispersed in both categories. 
We log-bin the pairs by separation and calculate the median D/T value for each $\Delta r$ bin, marked by the solid black curve. The shaded region represents the area between the 25th and 75th percentiles, showing dispersion around the median. The median curve for all dual AGN remains consistent around D/T values of $\sim 0.4 - 0.6$, with broad scatter suggesting xtremely weak dependence or minimal significance for this relationship. 
The median curves for ‘same-galaxy’ and ‘different-galaxy’ pairs, shown by green and orange curves, generally follow the overall trend, though there is a transition region $\Delta r \sim 5-7 \rm{kpc}$. This is expected as we are studying systems with different masses and properties at different stages of the merger, resulting in a mix of different scales.

In the upper panel, we show 1D distributions of galaxy separations. Most same-galaxy dual AGN have separations less than $5 \rm{kpc}$, peaking around $\Delta r = 3 \rm{kpc}$. Notably, at $5-7 \rm{kpc}$, both same-galaxy and different-galaxy dual AGN are present. At greater distances, most are ‘different-galaxy’ pairs, with a nearly flat distribution. 
Blue and red dashed lines show the number of the disk and bulge galaxies divided by the total number of galaxies within each $\Delta r$ bin, staying stable at around $0.6$ for disks and $0.4$ for bulges.However, there is a slight increase in $N_{\text{Disk}}/N_{\text{All}}$ to $0.75$ at the maximum separation distance of $30\rm{kpc}$, and this might be due to a low number of galaxies in that bin.

This suggests that different types of morphologies for Dual AGN are present independently of the pair distance. This holds whether the focus is on nearby pairs or those situated farther away, indicating that the morphologies are not subject to bias. 

\subsubsection{Morphologies as a function of SMBH parameters - Mass, Luminosity}

\begin{figure}
\centering
    \includegraphics[width=0.49\textwidth]{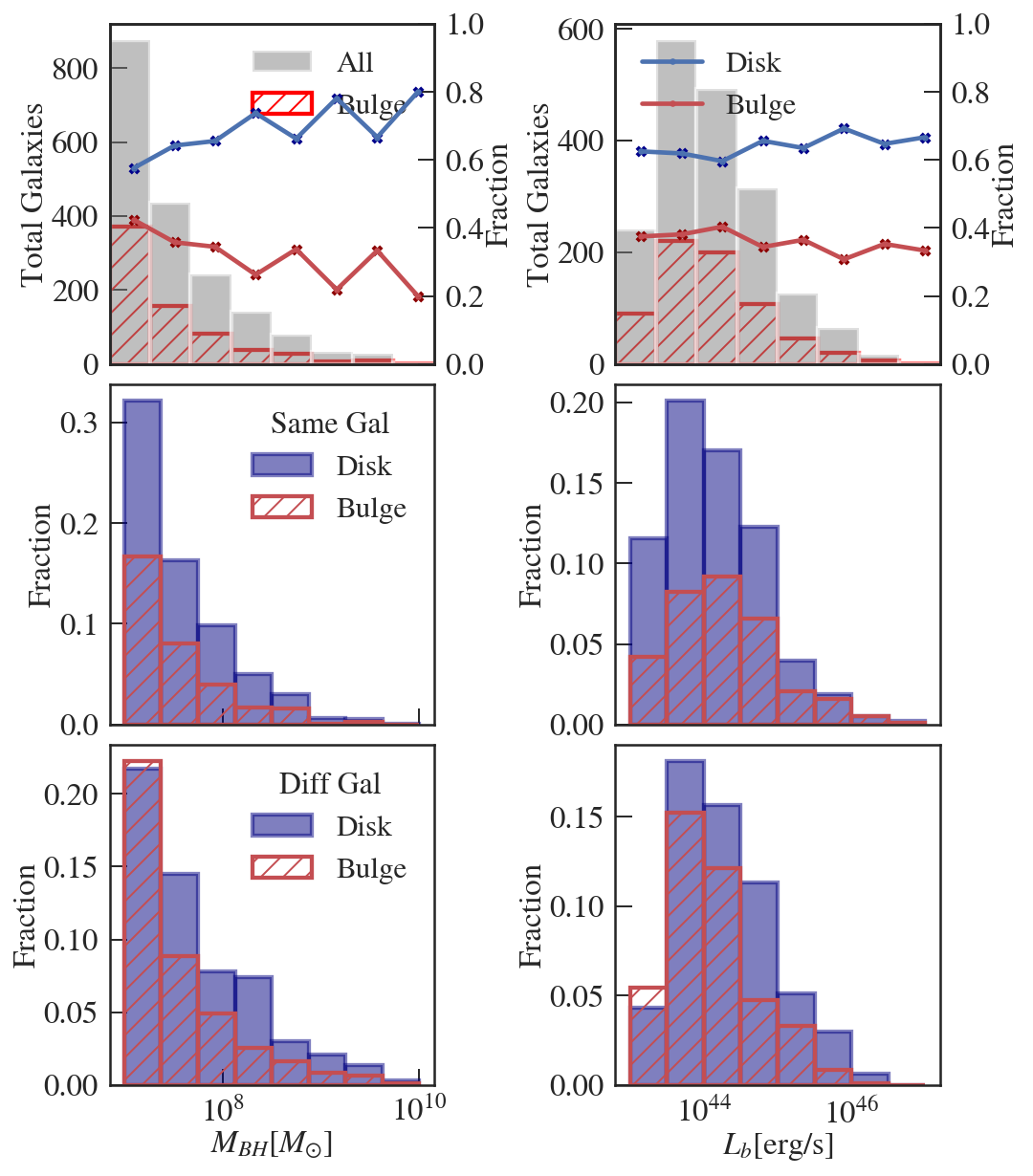}
  \caption{
  Distribution of disk and bulge galaxies for dual AGN systems according to black hole parameters: mass and luminosity. 
  The First row shows the distribution for all dual AGN systems (grey histogram). Here, the blue and red lines represent the fraction of disk and bulge galaxies within each parameter bin.
  The second/third row of the figure shows pairs of AGNs found within the same/different galaxy. In each subplot within these rows, disk/bulge galaxies are represented by blue/red histograms.
  }
  \label{disktopair}
\end{figure}

Figure \ref{disktopair} displays distributions of dual AGN parameters, specifically black hole mass and luminosity. The first column illustrates the SMBH mass distribution. The top panel encompasses the entire dual AGN population, represented by a grey histogram, with black holes hosted by bulge galaxies highlighted by the red dashed histogram. Moreover, we apply log-binning to the BH masses, calculating for each bin the fraction of disk to total galaxies (blue line) and bulge to total (red line). The top panel reveals a relatively steady fraction of disks, maintaining a value of approximately $0.6$ across the BH mass spectrum of $10^{7-10}M_\odot$.

We classify pairs into two types: the second row focuses on pairs in the same galaxy, while the third row shows those in separate galaxies, with dual AGNs in bulge and disk-dominated galaxies distinguished by red and blue histograms, respectively. In both categories, disk and bulge galaxies peak at the lowest BH mass of $10^7M_\odot$, declining for higher masses.

However, there is a notable difference between disk and spheroidal galaxies in the lowest mass bin. For same-galaxy pairs, disk galaxies outnumber spheroidal ones two-to-one, while the numbers are nearly equal for different-galaxy pairs.
This discrepancy primarily stems from the predominance of spheroidal galaxies hosting secondary black holes (BH2) in this mass bin. Specifically, $80\%$ of the spheroidal galaxies at the lowest mass bin host a secondary BH, with about $20\%$ of them at close distances ($\Delta r < 10$ kpc). This suggests that the higher number of spheroidal galaxies in the lowest mass bin is likely from minor mergers that may undergo significant tidal stripping, which is consistent with what we have highlighted in Figure \ref{q_DT}.

In the second column, luminosity distributions are shown. Total dual AGN distribution, represented by the grey histogram, peaks around $10^{44}\rm{erg/s}$ and declines at higher luminosities. A similar trend is also observed for spheroidal galaxies (red-striped histogram). The disk fraction remains steady at around $0.6$ across the luminosity range of $10^{43-47}\rm{erg/s}$.
Comparing same-galaxy pairs with different-galaxy pairs (2nd and 3rd rows), the trend is nearly identical, except at the lowest luminosities, where for same-galaxy pairs, disk galaxies outnumber bulges two-to-one, while spheroidal galaxies slightly prevail in different-galaxy pairs.
In the lowest luminosity bin for spheroidal galaxies, about $70\%$ are secondary BH hosts. Around $13\%$ of these are close-separation pairs ($\Delta r <10\text{kpc}$), suggesting the same explanation as for the lowest mass bin spheroidal abundance.

\subsection{Comparison with observations}
In the local universe, a significant portion of the stellar mass resides in disks \citep[according to][]{2009ApJ...696..411W, 2009MNRAS.393.1531G}. Aligning these observational results with galaxy formation simulations in a $\Lambda$CDM universe is challenging, as hierarchical assembly often results in systems dominated by bulges.

To better understand the effectiveness of the kinematic decomposition technique and compare it with observed data, in this section, we will fit the surface profiles to our sample of dual AGN, enabling a meaningful comparison between observational data and simulations. These surface profiles are fitted from mock observations based on the method described in \citet{2024arXiv240116608L}. We measure the S\'ersic index, effective radius, the Gini coefficient, and $M_{20}$ (the second-order moment of the region encompassing $20\%$ of the total flux) \citep{2003ApJ...588..218A, 2004AJ....128..163L}, for galaxies hosting our sample of dual AGN. In identifying late-stage mergers and classifying galaxies, $M_{20}$ has been shown to be more efficient than concentration and asymmetry \citep{2004AJ....128..163L}. Examining the relationship between the D/T parameter, Se\'rsic index, and effective radius can provide a better comparison of dual AGN morphology in observations and simulations.

\subsubsection{Morphology measures of mock observations}
 \begin{figure*}
\centering
  \includegraphics[width=0.99\textwidth]{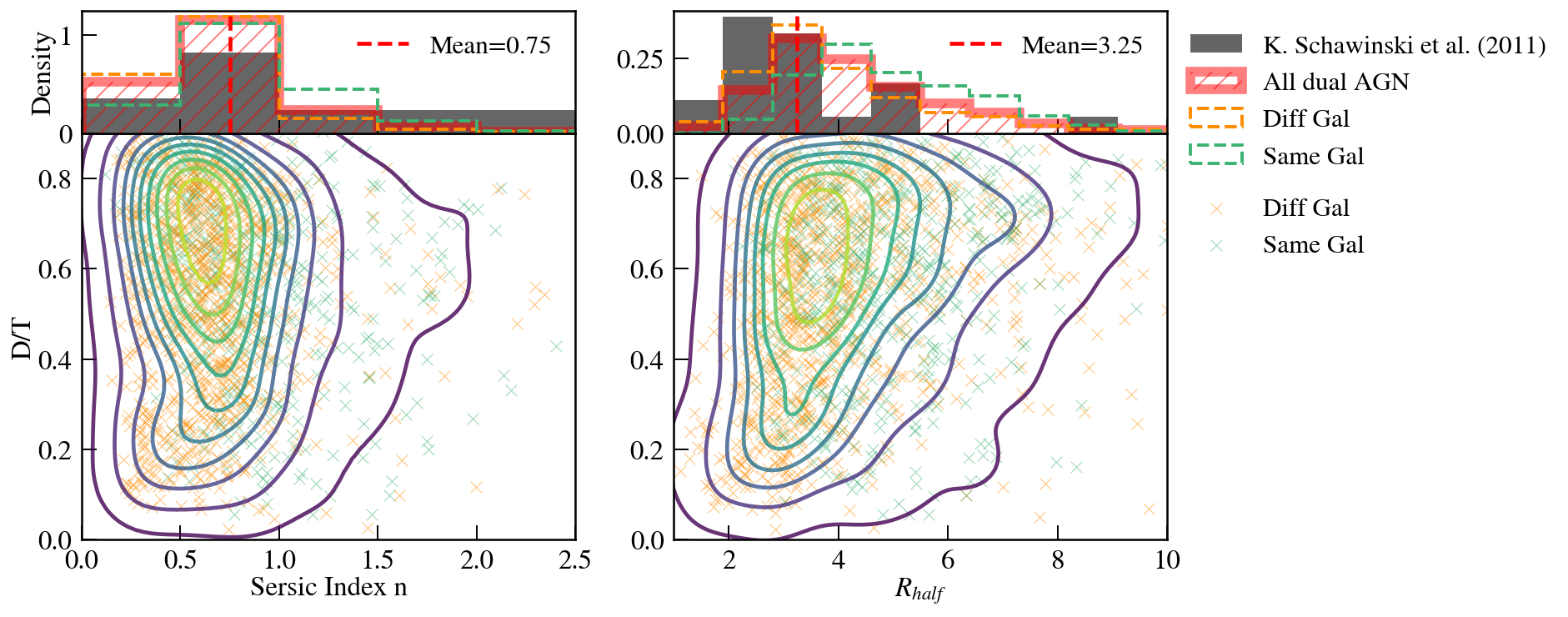}
  \caption{The dependence of the disk-to-total ratio (D/T) on the Se\'rsic index (bottom left panel) and effective radius (bottom right panel) for dual AGN. Green/orange crosses mark different/same galaxy pairs. The density contours are included for both. The top left and top right panels display the 1D distribution of the Se\'rsic index and effective radii, respectively. The red-striped histograms illustrate all dual AGNs, while the orange and green dashed lines represent the distributions for different and same galaxy pairs, respectively. The X-ray-selected AGN host galaxies at $z \sim 2$ from 
  \citet{2011ApJ...727L..31S} are shown by a solid black histogram.}
  \label{sersic}
\end{figure*}

Dual AGN host galaxies exhibit highly disturbed morphologies during the peak growth period at $z\sim2$. In such systems, surface profile fits can be challenging to interpret, and in some cases, the host galaxy may not be resolved \cite{2009ApJ...705..639B}. Around $3\%$ of our selected dual AGN host galaxies fall into this unresolved category.

In figure \ref{sersic}, we present the dependence of D/T on the S\'ersic index (bottom left). The majority of galaxies ($\sim83\pm2.4\%$) exhibit disk-like light profiles with $n < 1$ (Note that we limit our display to a Se\'rsic index below $2.5$, as it accounts for all but $1\%$ of the sample). This agrees with the results from HST WFC3/IR surveys conducted by \cite{2011ApJ...727L..31S}, affirming that a large portion of $z\sim2$ AGN host galaxies ($\sim80\%$) exhibit disk-like light distributions, with only a small fraction showing bulge-dominant profiles. 

The Se\'rsic index distribution for our entire dual AGN sample is depicted in the top-left panel, peaking at $0.75$ with a median of approximately $0.7$, as illustrated by the red-striped histogram. In contrast, the solid black histograms represent the results from \cite{2011ApJ...727L..31S}, indicating mean and median Se\'rsic indices of $2.54$ and $1.08$, respectively. Although the median values are reasonably consistent, there is a discrepancy in the mean values. This discrepancy, particularly the elevated $n$ values in their findings, could be attributed to their use of AGN-centric fits. Our analysis does not consider such fits since the AGN light is excluded from our image creation process.
Moreover, decoupling the host emission from AGN is challenging at $z\sim2$, and certain systematic biases are present, as noted by \cite{2008ApJS..179..283K}.
The distributions of different (orange dashed lines) and the same (green dashed lines) galaxy pairs display similar peaks.

Among galaxies with a S\'ersic index less than $1$, around $60.7\pm2.2\%$ are found to be disk-dominated galaxies according to the kinematic decomposition, having a D/T value greater than $0.5$.
The density contour plot reveals that dual AGN host galaxies with smaller Se\'rsic indices tend to have higher D/T values. The highest concentration is observed at a Se\'rsic index of $0.75$ and a D/T value of $0.7$.

\cite{2004ARA&A..42..603K} proposed a relationship between surface brightness and bulge types, distinguishing 'pseudo-bulges' from 'classical bulges'. This hypothesis was examined further by \cite{2008AJ....136..773F}, who discovered that a significant $90\%$ of pseudo-bulges possess a Se\'rsic index under $2$, a stark contrast to classical bulges, all of which have indices above $2$. Additionally, the Se\'rsic index in pseudo-bulges shows no association with factors such as luminosity, size, or their proportion in the galaxy, differing from classical bulges. This characteristic of having lower Se\'rsic indices for  pseudobulges is further elaborated in the studies \citep[e.g][]{2008AJ....136..773F, 2010ApJ...716..942F, 2009MNRAS.393.1531G}. This distinction further validates our analysis to classify pseudo-bulges as part of the disk components.

The distribution of effective radii (top right) for all dual AGN host galaxies is shown in the red striped histogram, with mean and median values of $\sim 3.5$ kpc and $\sim 4$ kpc, respectively. When segregating the sample based on different and same galaxy pairs, the median effective radii remain relatively consistent with all pairs in both cases, showing a peak at $\sim 3$ kpc and $\sim 4.5$ kpc, respectively. These values approximately align with observational data. \cite{2011ApJ...727L..31S} have reported mean and median effective radii at redshift $2$ to be $3.16$ kpc and $3.04$ kpc, respectively, as shown in the solid black histogram. In a related study, \cite{2016A&A...593A..22R} found that galaxies with $2 < z < 2.5$ possess an average effective radius ($r_{\text{eff}}$) of $1.8 \pm 0.1$ kpc, as measured by GALFIT-based techniques. The difference between the findings of these two investigations could be due to the latter not accounting for the surface brightness dimming effect, which leads to a strong correlation with redshift.
In the bottom right panel, the density contour plot is shown, with the highest concentration around $3$.

\begin{figure}
\centering
  \includegraphics[width=0.49\textwidth]{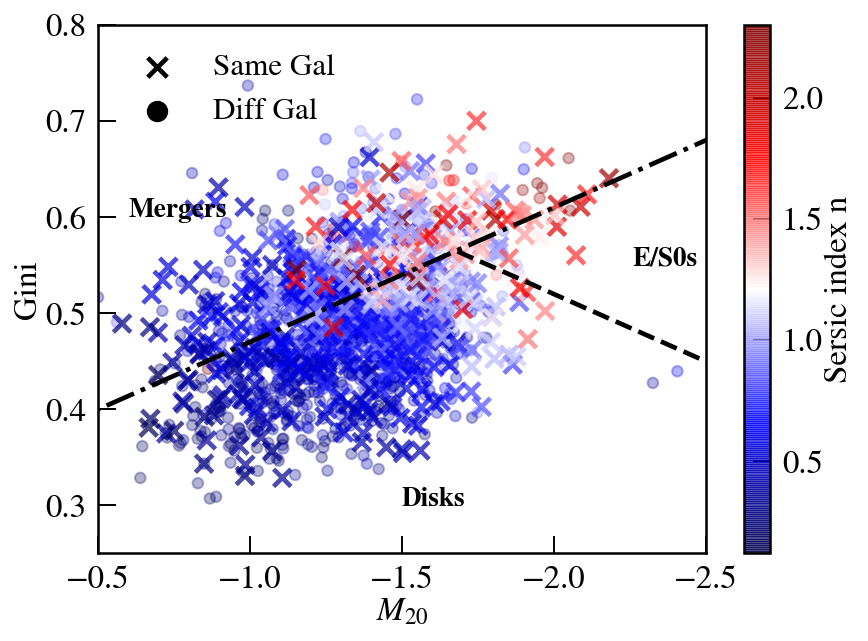}
  \caption{The Gini coefficient as a function of $M_{20}$ for all dual AGN color-coded by S\'ersic index. Dash-dotted lines indicating the boundaries between disk and elliptical galaxies and the dashed lines representing mergers, following \citet{2008ApJ...672..177L} methodology.}
  \label{gini}
\end{figure}

Figure \ref{gini} displays the $\text{Gini}-M_{20}$ relation for dual AGN host galaxies at $z\sim2$. The lines denote the boundary between mergers, disk galaxies, and ellipticals/S0s following \cite{2008ApJ...672..177L}. Around $34\%$ of galaxies lie above the merger line. The points are color-coded by the S\'ersic index, and most galaxies in the 'Disks' region have a S\'ersic index less than $1$. 
Dots represent pairs that reside in different galaxies, with $\sim33.4\%$ classified as mergers. In contrast, crosses denote pairs hosted by the same galaxy, with a notably higher percentage, around $42.8\%$, identified as mergers.

\subsubsection{Double quasars}

The study of close dual quasars throughout cosmic time is pivotal for comprehending galaxy mergers, particularly given their unique high-redshift and high-luminosity attributes \citep{2008ApJS..175..356H}. Despite advancements in identifying merging supermassive black holes at kpc-scales over the past decade \citep{2019NewAR..8601525D}, a considerable portion is found at low redshifts. The phase between $1 < z < 3$, a period marked by peak quasar activity and star formation, remains relatively untapped due to challenges posed by ground-based surveys. Ongoing studies like \cite{2022ApJ...925..162C} are progressively filling the gaps in the high-redshift, close-separation subspace of the dual AGN population.

While a number of dual active galactic nuclei have been identified in low-redshift mergers, there is a noticeable scarcity of dual disk quasars at $z > 2$. \cite{2023Natur.616...45C} recently revealed the first kiloparsec-scale ($3.48\rm{kpc}$) dual quasar within a galaxy merger occurring during the cosmic noon - J0749+2255. These quasars exhibit estimated bolometric luminosities of approximately $10^{46.59\pm0.03}$ and $10^{45.72\pm0.17}\rm{erg/s}$ powered by SMBHs with masses of $10^{9.07\pm0.4} M_\odot$, and $10^{9.12\pm0.4} M_\odot$. This high-redshift dual quasar is found within two massive galaxies, each with stellar masses of approximately  $10^{11.46\pm0.02} M_\odot$ and $10^{11.50\pm0.04} M_\odot$. 
Recent JWST NIRSpec IFU and MIRI IFU observations have revealed that J0749, initially identified as a dual quasar within a merging system, surprisingly hosts a massive rotating disk aligned perpendicularly to its two nuclei, contrasting with the expected irregular merger morphologies \citep{2024arXiv240308098I, 2024arXiv240304002C}.

\begin{figure}
\centering
\includegraphics[width=0.49\textwidth]{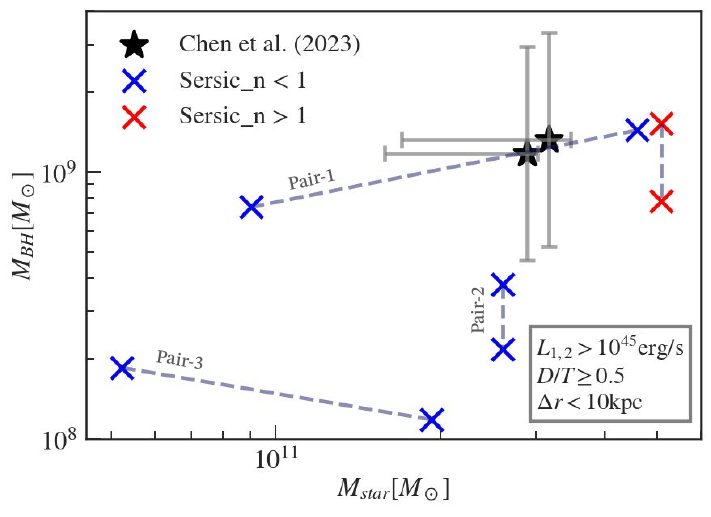}
\caption{ Black hole mass versus host galaxy stellar mass for the simulated dual AGN, with $L_{1,2}>10^{45}\rm{erg/s}$, $D/T\geq0.5$, and $\Delta r<10\rm{kpc}$. Dashed lines connect pairs. Blue and red crosses mark galaxies with s\'ersic indices less and greater than 1, respectively.
The black star markers indicate the disk-disk galaxies identified in a study by \citet{2023Natur.616...45C}, where error bars show the $1\sigma$  uncertainty.}
\label{disk_disk}
\end{figure}

To compare new observational findings with our simulated data, we identified dual AGN with close separation ($\Delta r<10 \rm{kpc}$), where both SMBHs have luminosities exceeding $10^{45}\rm{erg/s}$, and their host galaxies are categorized as disk galaxies based on the kinematic decomposition method \texttt{Mordor}, each with $D/T\geq0.5$. To align more closely with the latest observational data, we focused on higher-mass SMBH pairs ($M_{\text{BH},12}>10^8M_\odot$) and host galaxies with stellar masses exceeding $M_{*,12}>5\times10^{10}M_\odot$. This led to a final selection of four pairs, depicted in Figure \ref{disk_disk}. Blue and red crosses in the figure denote galaxies with Se\'rsic indices less than or greater than 1, respectively, distinguishing galaxies with disk-like and bulge-like profiles.
The black stars on the diagram mark the dual quasars from \cite{2023Natur.616...45C}, with error bars representing $1\sigma$ uncertainty.

\begin{table*}
\caption{ Properties for the simulated disky, luminous, massive dual AGN pairs. Each pair (Pair-1, Pair-2, Pair-3, Red-Pair) has two rows that correspond to its primary (BH1) and secondary black holes (BH2). The columns, listed from left to right, specify the name of the pair, the black hole mass, the stellar mass, the luminosity, the separation distance between the pair, the disk-to-total ratio, and the Se\'rsic index.}
\centering
\begin{tabular}{p{0.1\linewidth}p{0.1\linewidth}p{0.1\linewidth}p{0.1\linewidth}p{0.1\linewidth}p{0.1\linewidth}p{0.1\linewidth}p{0.1\linewidth}}
\hline
 & $M_{\text{bh}}[M_\odot]$ & $M_{*}[M_\odot]$ & L[\rm{erg/s}] & $\Delta r[\rm{kpc}]$ & D/T & Sersic n  \\
\hline
\textbf{Pair-1} \\
\hline
BH1 & $10^{9.15}$ & $10^{11.66}$  & $10^{46.38}$ & $6.37$ & $0.61$ & $0.89$  \\
BH2 & $10^{8.87}$ & $10^{10.96}$  & $10^{46.01}$ & $6.37$ & $0.59$ & $0.46$  \\
\hline
\textbf{Pair-2} \\
\hline
BH1 & $10^{8.58}$ & $10^{11.42}$ & $10^{45.54}$ & $0.99$ & $0.59$ & $0.56$ \\
BH2 & $10^{8.33}$ & $10^{11.42}$ & $10^{46.41}$ & $0.99$ & $0.59$ & $0.56$ \\
\hline
\textbf{Pair-3} \\
\hline
BH1 & $10^{8.27}$ & $10^{10.72}$  & $10^{45.23}$ & $5.16$ & $0.52$ & $0.29$  \\
BH2 & $10^{8.07}$ & $10^{11.29}$  & $10^{45.20}$ & $5.16$ & $0.64$ & $0.37$ \\
\hline
\textbf{Red-Pair} \\
\hline
BH1 & $10^{9.52}$ & $10^{11.08}$ & $10^{46.87}$ & $0.81$ & $0.51$ & $1.35$ \\
BH2 & $10^{8.76}$ & $10^{11.08}$ & $10^{46.17}$ & $0.81$ & $0.51$ & $1.35$ \\
\hline
\end{tabular}
\label{table}
\end{table*}

We find two SMBH pairs (pair-1 and pair-3) hosted by different galaxies and a third pair (pair-2) residing within a single host galaxy, where SMBHs are hosted by galaxies with $D/T\geq0.5$ indicating disk-dominated galaxies from kinematic decomposition and Se'rsic indices $n<1$ showing disk-like profiles. Notably, pair-1, with its larger stellar masses of $\sim10^{11.66}M_\odot$ for the primary and $\sim10^{10.96}M_\odot$ for the secondary BH, and higher luminosities with $\sim10^{46.36}\rm{erg/s}$ and $\sim10^{46.01}\rm{erg/s}$, aligns more closely with observational data. Furthermore, Pair-2 is also noteworthy, featuring a luminous ($L_{1}\approx10^{45.54}\rm{erg/s}$ and $L_{2}\approx10^{46.41}\rm{erg/s}$) and massive ($M_{\text{BH1}}\approx10^{8.58}M_\odot$ and $M_{\text{BH2}}\approx10^{8.33}M_\odot$) SMBH pair hosted by a single galaxy with a stellar mass of $\sim 10^{11.42}M_\odot$.
It is also important to note 'Red-Pair' from Figure \ref{disk_disk}, which has SMBH masses that are closest to those of the observed dual quasar, with values of $\sim 10^{9.52}$ and $10^{8.76}M_\odot$. However, the galaxy hosting this pair has a bulge-like profile with a Se\'rsic index of $n=1.35$ and a $D/T$ value of $0.51$ that is very close to the threshold. We will investigate this pair closely later.

\begin{figure}
\centering
\includegraphics[width=0.49\textwidth]{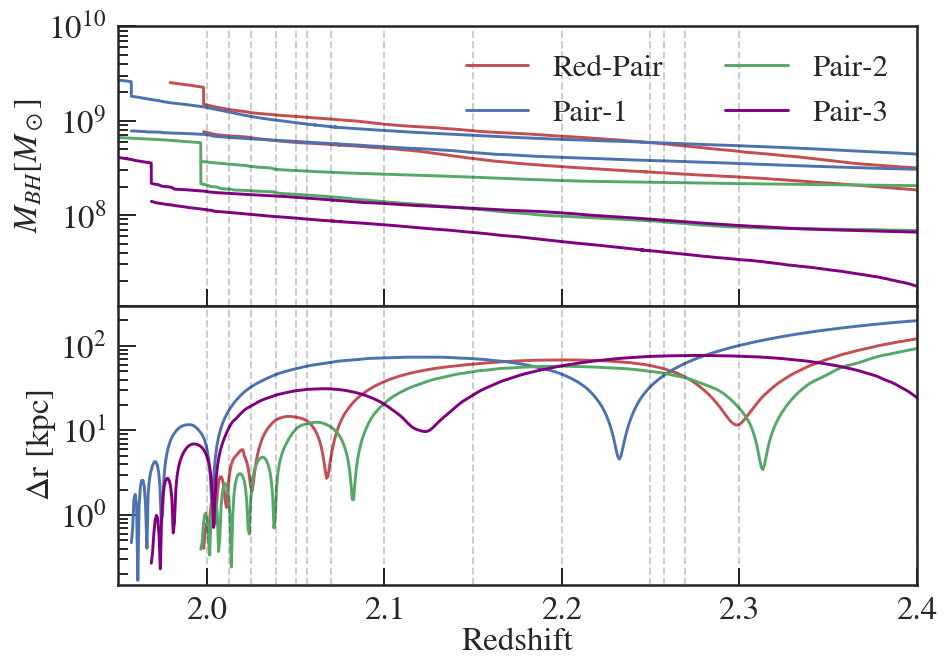}
\caption{The evolution of four selected pairs: Pair-1 (blue), Pair-2 (green), Pair-3 (purple), Red-pair (red), where we traced the SMBH masses (\textbf{top}, where two same-color lines represent two black holes in the pair) and the pair separation distance (\textbf{bottom}). We mark snapshots with vertical dashed lines.}
\label{evolution}
\end{figure}

The properties of these selected four pairs are detailed in Table \ref{table}, and Figure \ref{evolution} displays their SMBH masses (top) and pair separation distances (bottom) as a function of redshift from $z=1.95$ to $z=2.4$. From the evolution plot, we see that 'Red-pair' and 'Pair-2' have recently merged into one galaxy, while 'Pair-1' and 'Pair-3' are still in separate galaxies at Redshift 2 and are close to merging.

\begin{figure}
\centering
\includegraphics[width=0.49\textwidth]{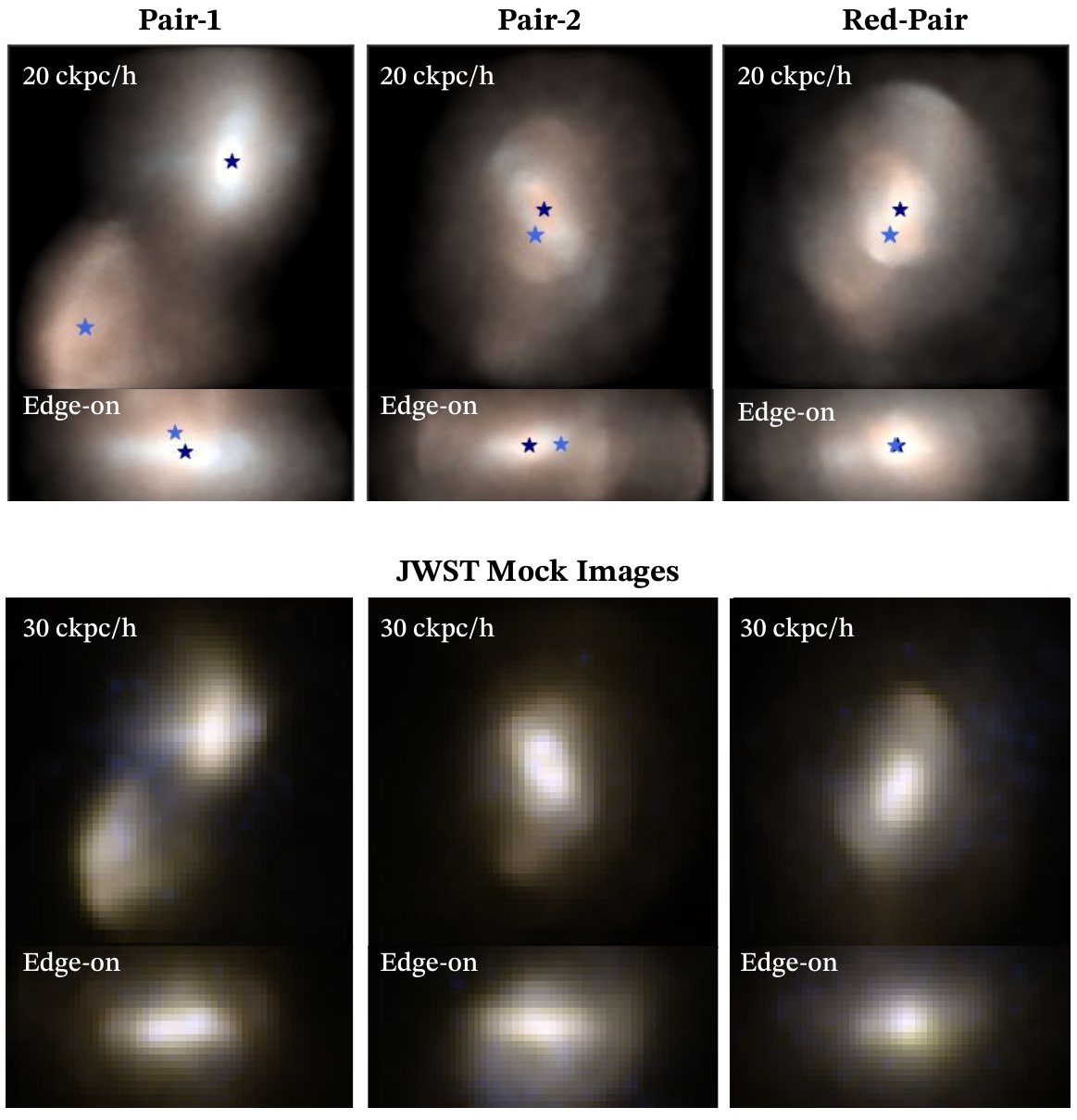}
\caption{
Stellar surface density maps for pairs in \ref{disk_disk}. 
The first, second, and third columns represent 'Pair-1', 'Pair-2', and 'Red-Pair' as listed in Table 1. Dark blue stars indicate primary black holes in pairs, while light blue stars represent secondary black holes. The first and the second rows show face-on and edge-on projections of simulated dual AGN, respectively. The bottom two rows show the same projections for mock images.}
\label{JWST}
\end{figure}

The first two rows of figure \ref{JWST} show the face-on and edge-on projections of the stellar surface density maps for the galaxies hosting Pair-1, Pair-2, and Red-Pair. The first column shows dual AGN hosted by different galaxies (Pair-1), while the second and third column depicts the dual AGN residing within a single galaxy (Pair-2, Red-Pair). The black holes in Pair-1 are separated by $\sim6.4\rm{kpc}$, while those in Pair-2 and Red-Pair are $\sim1\rm{kpc}$ and $\sim0.8\rm{kpc}$ apart, respectively. Additionally, the galaxies in the first two pairs exhibit disk-like profiles, as evidenced by their Se\'rsic indices of $0.89$ and $0.46$ for the host galaxies in Pair-1 and $0.59$ for the host galaxy of pair-2. The third galaxy shows a bulge-like profile, with a Se\'rsic index of $1.35$. 

The bottom two rows of figure \ref{JWST} show JWST mock images of these two systems. These images are made using a similar process as the one outlined in section \ref{sec: mock observations} but with filters and resolution chosen to emulate JWST NIRCam observations. Specifically, a drizzled resolution of 0.030" on the side and the NIRCam filters F356W, F277W, and F150W are used to create three mock observations of each system, which are combined to make the false color images present in the figure.

\begin{figure}
\centering
\includegraphics[width=0.49\textwidth]{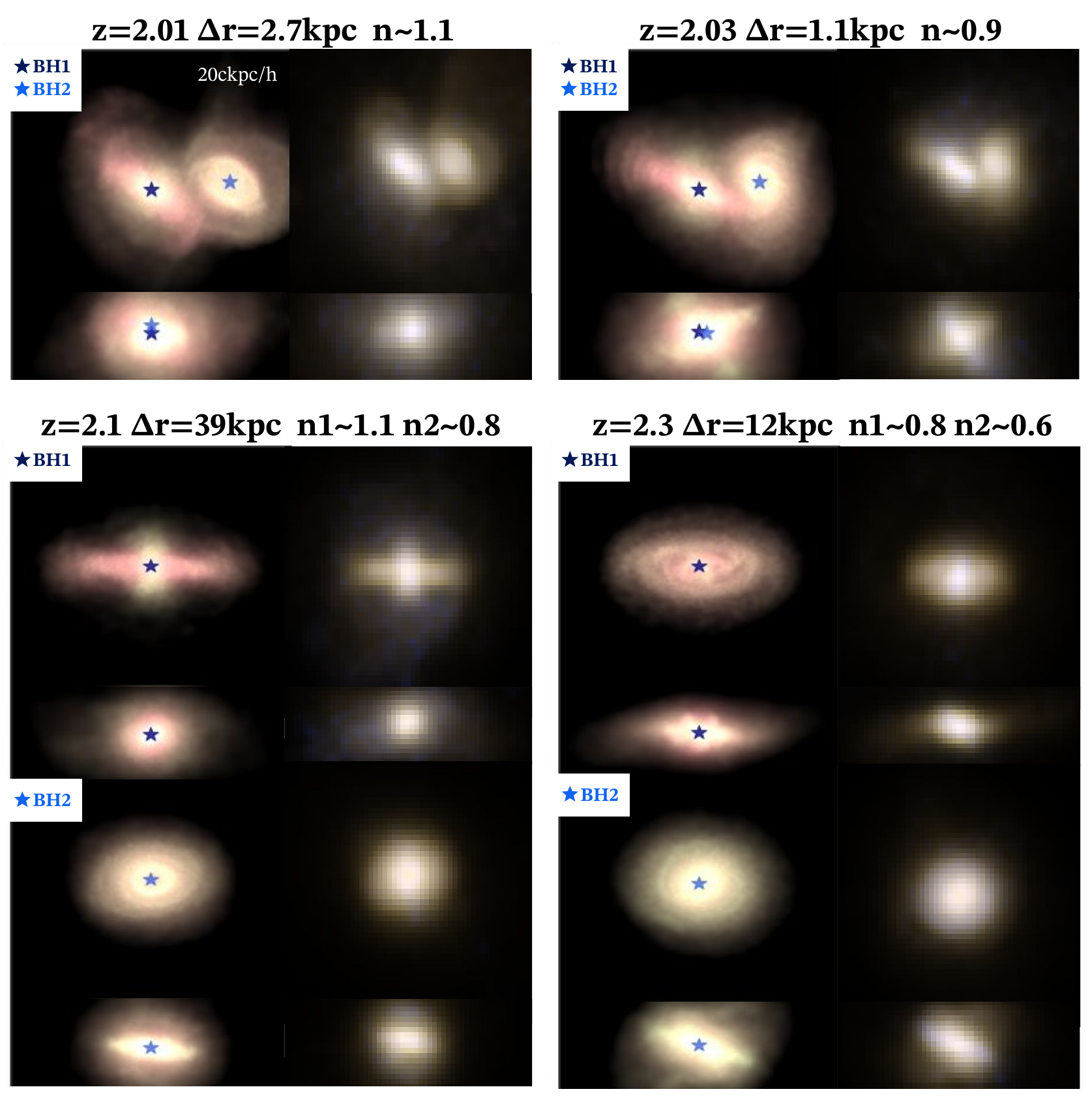}
\caption{Morphological Evolution of ‘Red-Pair’ Host Galaxies at four distinct redshifts: $z=2.01$, $z=2.03$, $z=2.1$, and $z=2.3$, color-coded by stellar age, with warmer colors indicating older stars. Each panel displays snapshots from these epochs, organized as follows: the first column shows the stellar density maps and the second column presents JWST mock images. The first and second rows provide face-on and edge-on views, respectively. In the top two panels, the black holes are within the same galaxy; thus, only one galaxy is depicted. In contrast, the bottom two panels depict the black holes in separate galaxies; the upper two sub-panels show the primary black hole (BH1) host galaxy from face-on and edge-on perspectives, while the lower two sub-panels display the secondary black hole (BH2) host galaxy in similar views. Redshifts, separation distances, and Se\'rsic indices are labeled on each panel.}
\label{red-pair}
\end{figure}

As mentioned earlier, 'Red-Pair' properties ($M_{bh1,2}$, $M_{\text{stellar}}$, $L$) closely resemble those of the observed dual quasar. However, they show a bulge-like profile according to its Se\'rsic index. Notably, the pair separation distance at $z=2$ is quite small ($\sim0.8\rm{kpc}$), and it is at the peak of the merging phase, as illustrated in Figure \ref{evolution}. 
Figure \ref{red-pair} shows the evolution of the 'Red-Pair' at different stages of galaxy merging. At $z=2.01$ and $z=2.03$, the pair is showing separation distances of $2.7\text{kpc}$ and $1.1\text{kpc}$, respectively, with corresponding Se\'rsic indices of $n\sim1.1$ and $n\sim0.9$. 
When we look back at 'Red-Pair' to the earlier redshifts of $z=2.1$ and $z=2.3$, we observe significant increases in the separation distances to $39\text{kpc}$ and $12\text{kpc}$, respectively.
The layout for each snapshot displays the host galaxy of the primary black hole in the upper sub-panels and that of the secondary black hole in the lower sub-panels. The Se\'rsic indices evolve to $n1\sim1.1$ and $0.8$ for the primary, and $n2\sim0.8$ and $0.6$ for the secondary, indicating a progression in their galactic structures. Additionally, using kinematic decomposition, the $D/T$ values for these redshifts change from approximately $0.37$ to $0.61$ for BH1, while BH2 remains nearly consistent at around $0.52$. These results align with the mock observational findings.

Comparing observational data with simulation results is crucial for better understanding the universe. Our simulation findings support the observational detection of a kpc-scale dual quasar in a galaxy merger during the cosmic noon period. The similarity between the two validates the parameters used in our model and strengthens our confidence in using such simulations to predict and understand the complex dynamics of galaxy mergers. 
These results emphasize the importance of disk-dominated galaxies hosting dual AGNs and provide a valuable reference for future observational studies, particularly with advanced IR facilities like the Nancy Grace Roman Space Telescope and the James Webb Space Telescope. These telescopes will enable detailed studies of quasar hosts at high redshifts, focusing on their morphological structures, thereby enhancing our detection and understanding of such systems across the universe.

\section{Conclusion / Discussion}

In this work, we examine the morphology of dual AGN (both active supermassive black holes) host galaxies at $z=2$ within the \texttt{Astrid} simulation. We focus on close separation ($\Delta r < 30 \rm{kpc}$) SMBH pairs that are both massive, with $M_{\text{BH,12}} > 10^7 M_\odot$, and luminous, with $L_{\text{bol,12}} > 10^{43} \rm{erg/s}$, following \cite{Chen2022b} definition. We identify $\sim 1000$ dual AGNs, with host galaxies having a stellar mass ranging from $10^{9-12} M_\odot$.

We apply \texttt{Mordor}, an improved galaxy kinematic decomposition algorithm, to analyze the morphology of selected dual AGN host galaxies. This algorithm segregates star particles based on their binding energy and circularity parameter, enabling us to separate five distinct galactic components - the thin and thick discs, pseudo-bulge, bulge, and halo.
To quantify the prominence of the disk structures, we introduce the disk-to-total ratio parameter as $D/T=M_{\text{Disk}}/M_{*}$, where $M_{\text{Disk}}=M_{\text{Thin Disk}}+M_{\text{Thick Disk}}+M_{\text{Pseudo-Bulge}}$. We classify galaxies into two categories: those dominated by disk structures as $D/T\geq 0.5$ and bulge structures as $D/T<0.5$, and we study their characteristics.

We calculate the mass ratios for each of the five galactic components in each stellar mass bin. As the stellar mass of galaxies increases from $10^9$ to $10^{11.5} M_\odot$, disk mass fraction (comprising thin and thick disks and pseudo-bulge) also increases from $17$ to $64\%$, while the total spheroidal (consisting of both bulge and halo) mass fraction decreases correspondingly. We find that the pseudo-bulge is the primary component within the disk mass composition, followed by the thin disk, while the thick disk has a relatively minor presence. The dominance of pseudo-bulge in massive dual AGN systems could be a signature that it has been a recent merger. On the other hand, the bulge is the dominant component for the combined spheroidal mass.

The distribution of stellar mass for disk-dominated galaxies peaks at around $10^{10.85}M_\odot$, while for bulge-dominated galaxies, it is slightly less at $\sim 10^{10.45}M_\odot$. The ratio of disk and bulge galaxies in each stellar mass bin for dual AGN host galaxies increases as we move to higher stellar mass bins, reaching a peak of about $3.8\pm1.5$ at around $10^{11.7}M_\odot$. In contrast, for galaxies hosting single AGN, the number ratio increases until it reaches its peak value of $2.3$ at a stellar mass of $10^{11}M_\odot$, and then decreases as we move to higher stellar mass bins, opposite to the trend observed for dual AGN host galaxies. In larger systems, galaxies hosting dual AGNs display a stronger tendency towards disk-like features and pseudo-bulge formation compared to those with single AGNs, which is expected since a pseudo-bulge usually indicates a recent merger, and dual AGNs, on average, are often found nearer to these merger events than single AGNs.

Within our sample of SMBH pairs, a significant majority, $\sim 61\%$, reside in disk-dominated galaxies characterized by $D/T>0.5$. The distribution of $D/T$ prominently peaks at approximately $0.7$. The same trend is observed for the single AGN host galaxies.
Among the pairs, the most prevalent scenario, accounting for $35\%$, is when both SMBHs in a couple belong to disk-dominated galaxies. Conversely, the least frequent configuration ($\sim 19\%$) is when bulge-dominated galaxies host both SMBHs. The remaining pairs show a mix, with each SMBH in a pair hosted by galaxies of differing morphologies. 

We observe no clear dependence among the duals between the pair separation distance and their host galaxy morphology. The distribution of pairs is scattered, with a median $D/T$ value ranging between $0.4$ and $0.6$ for all $\Delta r$, suggesting a lack of morphological preference based on pair separation. Moreover, the fraction of disk galaxies remains approximately constant at $0.6$ as black hole mass or AGN luminosity increases.

We compare our kinematic decomposition findings to photometrically decomposed AGN host galaxies at $z\sim 2$ from HST, as presented in \cite{2011ApJ...727L..31S}. The observed Se\'rsic indices and half mass radii align well, with our sample showing peaks at $0.75$ and $3.25\rm{kpc}$, respectively. Notably, approximately $83\pm2.4\%$ of our sample exhibits disk-like profiles with $n<1$, and among them, around $60.7\pm2.2\%$ are classified as disk-dominated based on kinematic decomposition. Additionally, the $\text{Gini}-M_{20}$ relationship indicates that about $34\%$ of dual AGN host galaxies sit above the merger line as per \cite{2008ApJ...672..177L}. It is worth noting that this cited study does not precisely target $z=2$, which may introduce discrepancies. As expected, Galaxies in the elliptical region predominantly exhibit higher Se\'rsic index values.

A recent discovery by \cite{2023Natur.616...45C} has revealed the first dual kiloparsec-scale ($3.48\rm{kpc}$) quasar within a galaxy merger at cosmic noon, featuring two supermassive black holes with significant bolometric luminosities and massive host galaxies that have disk-like profiles. JWST NIRSpec and MIRI IFU follow-up observations revealed a massive rotating disk perpendicular to its two nuclei, challenging the typical irregular merger morphologies \citep{2024arXiv240308098I, 2024arXiv240304002C}.
To check if our simulated data aligns with these new observational insights, in this study, we identify two dual AGNs with disk-like profiles as indicated by the Se\'rsic indices from their mock images that closely resemble the characteristics of the observed pair. These chosen dual AGNs exhibit similarities with the observed dual quasar in terms of the separation distance between the SMBHs ($\Delta r<10\rm{kpc}$), their mass ($M_{BH, 12}> 10^{8.5}M_\odot$), luminosity ($L_{\text{bol,12}}>10^{45.5}\rm{erg/s}$), and host galaxy characteristics such as stellar mass ($M_{*, 12}>10^{10.5}M_\odot$). We find that host galaxies of these SMBH pairs are disk-dominated according to our kinematic decomposition. 
Furthermore, we delved into the ‘Red-Pair,’ which most closely resembles the observed dual quasar, albeit with a bulge-like profile according to its Se\'rsic index. We tracked its morphological evolution through earlier snapshots, providing insights into how its morphology changes during different stages of the merging process.

The alignment between our simulated outcomes and observed data affirms the accuracy of our model's parameters. This strengthens our confidence in using these simulations for a clearer insight into galaxy mergers' complex morphology and dynamics.


\bibliographystyle{mnras}
\bibliography{main.bib}

\end{document}